\documentclass[floatfix,twocolumn,preprintnumbers,amsmath,amssymb,superscriptaddress]{revtex4-2}
\usepackage{amssymb}
\usepackage{amsmath}
\usepackage{makecell}
\usepackage{color}
\usepackage[pdftex]{graphicx}
\usepackage{multirow}
\usepackage{float}
\usepackage{mathtools}
\usepackage{comment}
\usepackage{colortbl}
\usepackage{nicematrix}
\usepackage[toc,page]{appendix}

\usepackage{makecell}
\usepackage{enumerate}
\usepackage{mathtools}
\usepackage{stmaryrd}
\usepackage{enumitem}
\usepackage{textcomp}
\usepackage{gensymb}
\usepackage{float}

\def \equi#1{\mathrel{\mathop{\kern 0pt\sim}\limits_{#1}}} 

\usepackage{diagbox}

\usepackage[overload]{empheq}
\usepackage{cases}

\newcommand{\leo}[1]{\textcolor{black}{ #1}}

\begin{document}

\title{From Maximum of Inter-visit Times to Starving Random Walks}

\author{L\'eo R\'egnier}
\address{Laboratoire de Physique Th\'eorique de la Mati\`ere Condens\'ee,
CNRS/Sorbonne University, 4 Place Jussieu, 75005 Paris, France}
\author{Maxim Dolgushev}
\address{Laboratoire de Physique Th\'eorique de la Mati\`ere Condens\'ee,
CNRS/Sorbonne University, 4 Place Jussieu, 75005 Paris, France}
\author{Olivier B\'enichou}
\address{Laboratoire de Physique Th\'eorique de la Mati\`ere Condens\'ee,
CNRS/Sorbonne University, 4 Place Jussieu, 75005 Paris, France}

\begin{abstract}

Very recently, a fundamental observable has been introduced and analyzed to quantify the exploration of random walks: the time $\tau_k$ required for a random walk to find a site that it never visited previously, when the walk has already visited $k$ distinct sites. Here, we tackle the natural issue of the statistics of $M_n$, the longest duration out of $\tau_0,\dots,\tau_{n-1}$. This problem belongs to the active field of extreme value statistics, with the difficulty that the random variables $\tau_k$ are both correlated and non-identically distributed. Beyond this fundamental aspect, we show that the asymptotic determination of the statistics of $M_n$ finds explicit applications in foraging theory and allows us to solve the open $d$-dimensional starving random walk problem,  in which each site of a lattice initially contains one food unit, consumed upon visit by the random walker,  which can travel $\mathcal{S}$ steps without food before starving. Processes of diverse nature, including regular diffusion, anomalous diffusion, and diffusion in disordered media and fractals, share common properties within the same universality classes.
\end{abstract}
\maketitle

The territory covered by random walks (RWs) constitutes a fundamental property with significant implications in quantifying the efficiency of diverse stochastic exploration processes, ranging from animal foraging behaviors~\cite{gordon1995development} to the trapping of diffusing molecules~\cite{hollander_weiss_1994}. Usually, this explored territory is quantified by the number $N(t)$ of distinct sites visited at time $t$~\cite{vineyard1963number}. Its average, variance and, in some cases, full distribution have been determined analytically \cite{Hughes_1995,Gall_1991,gillis_2003,biroli2022number}. However, since $N(t)$ is a cumulative quantity, it does not describe the detailed dynamics of the exploration process. In particular, it does not differentiate between trajectories in which new sites are discovered at an almost regular rate, and those in which they are essentially found towards the end of exploration, corresponding to long periods of time with no new sites discovered. 

\begin{figure}
    \centering
    \includegraphics[width=\columnwidth]{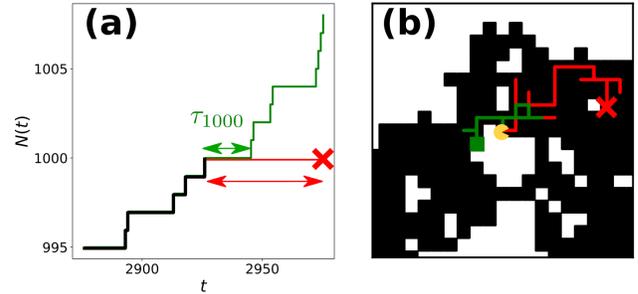}
   \caption{{\bf Inter-visit Times }{\bf (a)} The inter-visit times $\{\tau_k\}$ (horizontal steps), defined as the time intervals between increments of the number $N(t)$ of distinct sites visited, control the exploration process. After visiting $N(t)=1000$ sites the discovery of the $1001^{\rm st}$ new site can be either "short" (green arrow) or "long" (red arrow). Here, we determine the statistics of the maximum $M_n$ of $\{\tau_k\}_{k<n}$. 
   \\  {\bf The starving RW model}. {\bf (b)} Each site of a lattice initially contains one food unit, consumed upon visit by the RW, which can travel $\mathcal{S}$ steps without food
before starving.  Two sample trajectories (green and red) associated with the evolution scenarios of $N(t)$ in (a) are displayed. A forager (yellow) has eaten $N(t)=1000$ food units (the domain depleted of food is black, and  $\mathcal{S}=50$). Following the green trajectory, it  finds rapidly a new food unit (green square). On the red trajectory, it fails to find food before starving (at red cross).}
    \label{fig:Intro}
\end{figure}

Very recently, as a first step to account for this disparity between random trajectories,  another fundamental quantity was introduced \cite{regnier2022complete, regnier2023universal}: the time $\tau_k$ required for the RW to find a site that it never visited previously when $k$ distinct sites have already been visited, see Fig.~\ref{fig:Intro}. These random variables are indeed very useful because they encompass the full dynamics of the visitation statistics \cite{regnier2022complete,regnier2023range,regnier2023universal}. The knowledge of the statistics of a given $\tau_k$ variable is however insufficient to characterize the long periods of time with no new sites discovered, which can deeply impact the exploration process (see Fig.~\ref{fig:Intro}\textbf{a}).  

In this Letter, we provide a quantitative characterization of these long time periods  by determining the asymptotic statistics of the maximum  $M_n$ of the $\tau_k$,  $M_n={\rm max}(\tau_0,\dots,\tau_{n-1})$ \footnote{With the convention $N(0)=1$ and $\tau_0=0$, the starting site being visited at the beginning of the exploration process.}. This fundamental question belongs to the 
 domain of extreme value statistics (EVS), which has attracted considerable attention in recent years due to its connection  with the statistics of extreme events \cite{Fisher:1928,Gumbel:1958,MAJUMDAR2021,Vezzani2019}. Applications are found in fields as diverse as  disordered systems \cite{Derrida:1981,Derrida:1986}, random matrices \cite{Tracy:1994,Majumdar:2014} and search algorithms \cite{Majumdar:2000,ben:2001,Meyer:2021}. 
Here, the technical difficulty is that  the territory visited by the RW is incessantly updated. As a result, 
 the random variables $\tau_k$ are both correlated and non-identically distributed. Importantly,  these characteristics are not given a priori but are generated by the RW itself. 
 
Beyond this theoretical  aspect,  the  determination of the statistics of $M_n$   finds explicit applications in foraging theory, particularly in the context of the starving RW model \cite{sowinski2023semantic,benichou2014depletion,Benichou_2016,benichou2018frugal,le1991range,Randon2022}, which describes depletion-controlled starvation of a RW forager. In the original version of this model~\cite{benichou2014depletion}, the RW survives only if the time elapsed until a new food-containing site is visited is less than an intrinsic metabolic time $\mathcal{S}$ (see Fig. ~\ref{fig:Intro}\textbf{b}). Such a situation is commonly encountered at various scales \cite{klages2023search}, ranging from microscopic organisms, such as bacteria \cite{Passino:2012}, to larger creatures like insects, foraging mammals~\cite{orlando2020power} and robots \cite{winfield2009towards}. So far, the analytical results on starving RWs have essentially been limited to one dimension (resp. infinite dimension, corresponding to a  mean field solution), where it was found that: (i) the mean number $\langle N_\mathcal{S}\rangle$ of units of food collected at starvation is proportional to  $\mathcal{S}^{1/2}$ (resp. exponential in $\mathcal{S}$), (ii) the mean lifetime $\langle T_\mathcal{S}\rangle$ is proportional to $\mathcal{S}$ (resp. exponential in $\mathcal{S}$), and (iii) the distributions of these two observables admit a single-parameter scaling. Since the original model \cite{benichou2014depletion},  several  lines of extension have been considered, including resource renewal \cite{Chupeau2016Ressource}, long-range food detection \cite{sanhedrai2020lifetime,Sanhedrai2021}, penalties on long moves \cite{Krishnan2018}, switching on several modes of motion \cite{campos2021optimal}. The only studies in higher dimensions concern the mean-field approach mentioned above \cite{benichou2014depletion,Benichou_2016},  numerical simulations \cite{sanhedrai2020lifetime,Sanhedrai2021}  and a scaling result on the mean lifetime $\langle T_{\mathcal S}\rangle\propto {\mathcal S}^2$ in the particular case of the $2$D situation \cite{regnier2023universal}. 

Here, we provide analytical results for the $d$-dimensional starving nearest-neighbor RW model, which constitutes an open problem. 
Our starting point is the  observation that  the knowledge of the statistics of the maximum $M_n$ is a key step to describe the dynamics of a starving RW. 
Indeed, a  starving RW is still alive after $n$ units of food have been collected if (and only if) $M_n<\mathcal{S}$. 
In this Letter, we derive analytically the long-time asymptotic distribution of, first, $M_n$ and, second, key observables pertaining to starving RWs: the lifetime $T_\mathcal{S}$,  the number $N_\mathcal{S}$ of units of food collected
when starvation occurs, and the position $\vec{R}_\mathcal{S}$ of the walker when it starves.
A wide range of processes, including $d$-dimensional regular diffusion, anomalous diffusion, and diffusion  in disordered media and fractals, fall into the same universality classes.

\vspace{5pt}

{\it Maximum of the inter-visit times.---} 
We consider the general situation of a discrete-time symmetric Markovian RW on a lattice of fractal dimension $d_\text{f}$  ($d_\text{f}$ being equal to $d$ in the particular case of a $d$ dimensional Euclidean lattice). The RW dynamics is characterized by the walk dimension $d_\text{w}$ given by the typical displacement $ r(t)  \propto t^{1/d_\text{w}}$ after $t$ steps. 
Recurrent (shown \cite{Hughes_1995,ben2000} to be obtained for $\mu\equiv d_\text{f}/d_\text{w}<1$)  and marginal ($\mu=1$) RWs visit any site with  probability one, whereas transient ($\mu>1$) RWs have a finite probability not to visit it. We define by $\tau_k$ the time elapsed 
 between the visits to the $k^{\rm th}$ and the $(k+1)^{\rm st}$ distinct sites
\cite{regnier2023universal,benichou2014depletion} and by $M_n$ the maximum of the inter-visit times $\tau_k$, $M_n=\max(\tau_0,\ldots,\tau_{n-1})$. We report here that the rescaled random variable  
\begin{align}
    x_n=
\begin{cases}
M_n/\left\langle M_n\right\rangle, &  \mu \leq 1,\\
(M_n-\left\langle M_n\right\rangle)/\sqrt{\text{Var}( M_n) },                      & \mu>1,
\end{cases}\label{eq:x}
\end{align}
where the scalings with $n$ of the averages and standard deviations of $M_n$ are given by 
\begin{align}
    \left\langle M_n\right\rangle \propto 
\begin{cases}
n^{1/\mu} &  \mu<1\\
\sqrt{n}      & \mu=1\\
(\ln n)^{1/\mu+1}                       & \mu>1
\end{cases}\label{eq:AvM}
\end{align}
and 
\begin{align}
    \sqrt{\text{Var}( M_n) } \propto 
\begin{cases}
n^{1/\mu} &  \mu<1\\
\sqrt{n }      & \mu=1\\
(\ln n)^{1/\mu}                       & \mu>1,
\end{cases}\label{eq:VarM}
\end{align}
is asymptotically ($n\rightarrow\infty$) distributed according to a cumulative distribution function (CDF) $\Xi_\mu(x)$. Even if process-dependent, $\Xi_\mu(x)$ displays the following universal asymptotic behaviors, depending on the nature of exploration, recurrent, marginal or transient:

For recurrent RWs ($\mu<1$),
\begin{align}
    -\ln \Xi_\mu (x) \propto 
    \begin{cases}
    x^{-\mu} &\mbox{ for } x\ll 1 , \\
        E_1(Ax)\propto e^{-Ax}/x &\mbox{ for } x \gg 1, 
    \end{cases}
    \label{eq:RecDistMn}
\end{align}
where $E_1$ is the exponential integral function and $A$ is a process-dependent constant.

For marginal RWs (${\mu=1}$), the distribution obeys (up to log corrections)
\begin{align}
    -\ln \Xi_1 (x) \propto 
    \begin{cases}
        x^{-2} &\mbox{ for } x \ll 1 \\
        e^{-Bx^{1/2}} &\mbox{ for } x\gg 1 
    \end{cases}
    \label{eq:MargDistMn}
\end{align}
where $B$ is a process-dependent constant.

For transient RWs, $\Xi_\mu(x)$ does not depend on $\mu$ and is given by the celebrated Gumbel distribution \cite{MAJUMDAR2021}
\begin{align}
    -\ln \Xi_\infty (x)=\exp\left[-\pi x/\sqrt{6}-\gamma_E \right]  
    \label{eq:TransDistMn}
\end{align}
where $\gamma_E$ is the Euler constant.

Striking qualitative differences between  recurrent and transient RWs emerge: while for recurrent and marginal RWs the standard deviation of $M_n$ is always comparable to its mean value, this is not the case of  transient RWs for which the standard deviation of $M_n$ is negligible in comparison to its mean. As a consequence, $M_n$ becomes asymptotically deterministic in the latter case. Besides the average and variance, the asymptotic distribution of the rescaled maximum in the recurrent and marginal cases  is very different from the usual Gumbel distribution for random variables with stretched exponential tails distribution, as can be seen by comparing Eq.~\eqref{eq:RecDistMn} and \eqref{eq:MargDistMn} to Eq.~\eqref{eq:TransDistMn}. As shown below, this is the signature of  strong aging effects for $\mu \leq 1$. 

We now sketch the main steps involved in  obtaining these results (see SM for detailed derivations \footnote{See Supplemental Material which includes Refs.~\cite{renyi1953theory,Bouchaud:1990, majid1984exact,Kolmogorov1954,Barlow1988,bunde1990multifractal,jones1996transition,regnier2023record,bray2013persistence,redner_2001}}). We emphasize that the treatment of the recurrent, marginal and transient cases have to be differentiated due to the disparities in the visitation process, as described in \cite{regnier2023universal}. The calculations are based on the hypothesis that the events $\lbrace \tau_k < \mathcal{S} \rbrace$ (with $k\leq n-1$) are asymptotically ($n\to\infty$) effectively independent. This key hypothesis is extensively checked numerically in SM, and self-consistently analytically checked below. The effective independence of $\tau_k$ allows to express the CDF of $M_n$ via the CDF of $\tau_k$:
\begin{align}
    \mathbb{P}(M_n\leq T)&\approx \prod_{k=0}^{n-1}\mathbb{P}(\tau_k\leq T) \nonumber \\
    &\approx \exp\left[ -\sum_{k=0}^{n-1}\int_T^\infty F_k(\tau) {\rm d}\tau \right] , \label{eq:CDFmax}
\end{align}
where $F_k(\tau)$ is the probability distribution function of $\tau_k$, whose asymptotics were determined recently in \cite{regnier2023universal} (see also SM for refined characterization in the marginal case).

For recurrent walks ($\mu<1$), the probability distribution function of $\tau_k$ presents a scaling form, $F_k(\tau)=k^{-1-1/\mu}\psi(\tau/k^{1/\mu})$, where $\psi(u)$ is algebraic at small $u$ and exponential at large $u$~\cite{regnier2023universal}. This implies that the CDF of $M_n$ also has a scaling form, since
\begin{align}
    &- \ln \mathbb{P}(M_n\leq T)\approx  \sum_{k=0}^{n-1}\int_T^\infty k^{-1-1/\mu}\psi(\tau/k^{1/\mu}) {\rm d}\tau \nonumber \\
    &\approx \int_{0}^{n/T^\mu} \frac{{\rm d} v}{v}\int_{v^{-1/\mu}}^\infty \psi(u) {\rm d}u  =- \ln \Xi_\mu (T/n^{1/\mu}) 
    \label{eq:RecScale}.
\end{align}
This leads to the asymptotics of Eq.~\eqref{eq:RecDistMn}.

For marginal walks ($\mu=1$), the CDF of $M_n$ is shown to be dominated  by the behaviour of $F_k(\tau)$ in the regime $\sqrt{k}\sim \tau $ corresponding to the typical time needed to exit the largest fully visited spherical domain, determined in~\cite{Dembo2007,regnier2023universal}. 
Extending this approach to the determination of the scaling of the exit time of the next largest fully visited domains, we show in SM that $F_k(\tau)=k^{-3/2}\psi(\tau/\sqrt{k})$ with $\psi(u)\propto u^{-3}$ at small $u$ and $-\ln\psi(u)\propto \sqrt{u}$ at large $u$ (up to log corrections). In turn, this scaling form allows one to adapt the steps of Eq.~\eqref{eq:RecScale} to the marginal case. We obtain that the CDF of $M_n/\sqrt{n}$ has asymptotically a single scaling parameter ($\left\langle M_n \right\rangle \sim \sqrt{\text{Var}(M_n)}\sim \sqrt{n}$ up to log prefactors) and converges to the cumulative distribution~$\Xi_1$ of Eq.~\eqref{eq:MargDistMn}.

For transient walks ($\mu>1$), for times $\tau\ll k^{1+1/\mu}$, the probability distribution function of $\tau_k$ is independent of $k$ and stretched exponentially distributed of exponent $\frac{\mu}{1+\mu}$ \cite{regnier2023universal}. By showing that the CDF of $M_n$ is controlled by this early time regime of $F_k(\tau)$, we obtain that the limit distribution ($n\to\infty$) is the Gumbel law displayed in Eq.~\eqref{eq:TransDistMn}.

\begin{figure}
    \centering
    \includegraphics[width=\columnwidth]{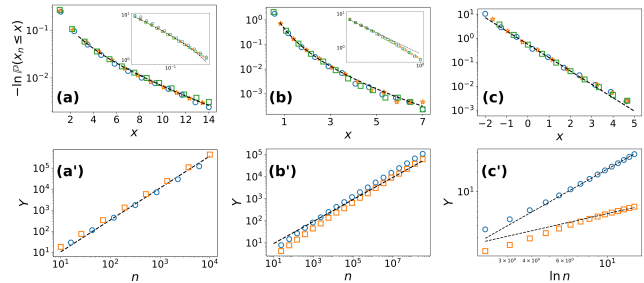}
    \caption{{\bf Maximum of inter-visit times.} {\bf (a)-(c)} $M_n$ CDF as a function of the rescaled variable $x_n$ defined in Eq.~\eqref{eq:x} (insets show them at small $x_n$ values) and {\bf (a$^\prime$)-(c$^\prime$)} the corresponding averages (blue circles) and standard deviations (orange squares) of $M_n$. The black dashed lines correspond to the best fit of Eqs.~\eqref{eq:AvM} to \eqref{eq:TransDistMn}. Different universality classes are represented by {\bf (a)} RWs on a percolation cluster, $\mu\approx 0.659$ (recurrent), $n=1389$, $3727$ and $10^4$;  {\bf (b)} nearest neighbour $2$D RWs, $\mu=1$ (marginal), $n\approx 2\cdot 10^7$, $10^8$ and $3\cdot 10^8$ \cite{first};
    {\bf (c)} nearest neighbour $3$D RWs, $\mu=3/2$ (transient), 
    $n\approx 4\cdot 10^4$, $2\cdot 10^5$ and $10^6$  \cite{second}. 
    Increasing values of $n$ are represented by blue circles, orange stars, and green squares.}
    \label{fig:Max}
\end{figure}

Finally, we provide a self-consistent analytical check of the effective independence of the $\{\tau_k\}$ used in Eq.~\eqref{eq:CDFmax}. This constitutes  an extension of the argument of Ref.~\cite{carpentier2001}, originally given for Gaussian correlated but identically distributed random variables.\leo{ The idea is that one can neglect the effect of the correlations on the statistics of the maximum if these correlations are typically much smaller than the maximum's fluctuations induced by the random variables without these correlations.}\\
\leo{To make this criteria quantitative, we consider the typical correlation in the $n$ random variables $\lbrace \tau_k \rbrace_{k<n}$, $\sqrt{\text{Cov}(\tau_{n/4},\tau_{3n/4})}$, and compare them to the maximum's standard deviation, $\sqrt{\text{Var}(M_n)}$, supposing that these $\tau_k$ are independent, which are given by \eqref{eq:VarM}. An upper bound of the inter visit times correlation is given by the Cauchy-Schwarz inequality, $\text{Cov}(\tau_{n/4},\tau_{3n/4}) \leq \sqrt{\text{Var}(\tau_{n/4})\text{Var}(\tau_{3n/4})}\propto \text{Var}(\tau_n)$, which is known from \cite{regnier2023universal} for any value of Markovian RW. For recurrent RWs, this leads to
\begin{align}
\text{Cov}(\tau_{n/4},\tau_{3n/4})\leq n^{2/\mu-1} \ll n^{2/\mu}\sim\text{Var}(M_n)
\end{align}
and for marginal RWs, to 
\begin{align}
    \text{Cov}(\tau_{n/4},\tau_{3n/4})\leq \sqrt{n} \ll n\sim\text{Var}(M_n).
\end{align}
For transient RWs the variance of the inter visit time is constant so that 
\begin{align}
    \text{Cov}(\tau_{n/4},\tau_{3n/4})\leq \text{cst.} \ll \left(\ln n\right)^{2/\mu}\sim\text{Var}(M_n).
\end{align}}
We conclude that, in all cases, the typical fluctuations dominate the typical cross-correlations for all RW classes so that $\lbrace \tau_k  \rbrace $ are effectively independent and hence Eq.~\eqref{eq:CDFmax} is self-consistently checked. \leo{Note that, contrary to the central limit theorem where long-range correlations can deeply impact the asymptotic law of the sum of $n$ random variables, the maximum is less sensitive to cross-correlations as its fluctuations are relatively (compared to the mean) larger. As an example, for $n$ i.i.d. random variables with finite variance, while the relative fluctuations of the sum decays as $1/\sqrt{n}$, the relative fluctuations of the maximum decay logarithmically as $1/\ln n$ \cite{MAJUMDAR2021}. }

Further validation of our results is provided in Fig.~\ref{fig:Max}, in which we numerically test Eqs.~\eqref{eq:AvM} to  \eqref{eq:TransDistMn} on representative recurrent, marginal and transient RW models. We see a very good agreement between our analytical predictions and numerical simulations. The diversity of these examples  demonstrates the broad applicability of our theoretical approach.

\vspace{5pt}

\textit{Starving Random Walks.---}
We now show that the knowledge of the CDF of $M_n$ is an essential tool to quantify the interplay between the amount of the resource consumed and the lifetime of a starving RW, as introduced above.
We first consider  the number of sites visited at starvation $N_\mathcal{S}$, which is a key observable to quantify the exploration efficiency of starving RWs \cite{benichou2014depletion}. At starvation, at least $n$ sites have been visited if and only if all the first $n$ times between two visits are strictly less than the metabolic time $\mathcal{S}$, ${\tau_0<\mathcal{S}},\dots,{\tau_{n-1}<\mathcal{S}}$. In other words,
\begin{align}
    \mathbb{P}(N_\mathcal{S} \geq n)&=\mathbb{P}(M_n< \mathcal{S})\label{dist_N_S},
\end{align}
so that the distribution of $N_\mathcal{S}$ is directly deduced from that of $M_n$. In particular, we obtain (see SM for derivation and numerical verification) the scaling with $\mathcal{S}$ of the first two cumulants of~$N_\mathcal{S}$,
\begin{align}
    \left\langle N_\mathcal{S}\right\rangle  ,\sqrt{\text{Var}(N_\mathcal{S})} \propto
\begin{cases}
\mathcal{S}^{\mu} &  \mu<1\\
\mathcal{S}^2      & \mu=1\\
\exp\left[ \left(\mathcal{S}/a\right)^{\mu/(1+\mu)}\right]                       & \mu>1
\end{cases}\label{eq:momN}
\end{align}
where $a$ is a positive constant. 

\begin{figure}[t!]
    \centering
    \includegraphics[width=\columnwidth]{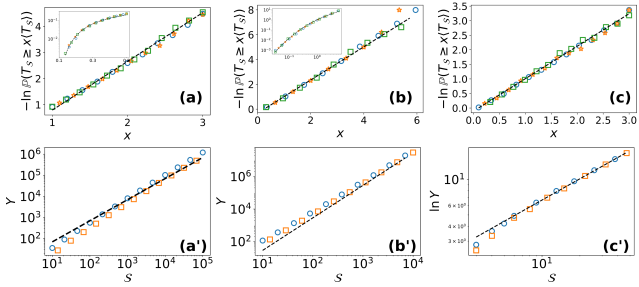}
    \caption{{\bf Lifetime of a starving RW.} {\bf (a)-(c)} $T_\mathcal{S}$ distributions as a function of the rescaled variable $x\equiv T_\mathcal{S}/\langle T_\mathcal{S}\rangle$ (the insets show the behavior at small $x$-values) and {\bf (a$^\prime$)-(c$^\prime$)} the corresponding averages (blue circles) and standard deviations (orange squares) of $T_\mathcal{S}$. The black dashed lines correspond to the best fit of the theory. Different universality classes are represented by {\bf (a)} RWs on a percolation cluster, $\mu\approx 0.659$ (recurrent), $\mathcal{S} = 14667$, $31622$ and $68129$;  {\bf (b)} nearest neighbour $2$D RWs, $\mu=1$ (marginal), $\mathcal{S}= 2335$, $4832$ and $10^4$; {\bf (c)} nearest neighbour $3$D RWs, $\mu=3/2$ (transient), $\mathcal{S}=12$, $17$ and $22$.  Increasing values of $\mathcal{S}$ are represented by blue circles, orange stars, and green squares.}
    \label{fig:Lifetime}
\end{figure}

Next, the distribution of $N_\mathcal{S}$ gives access to that of the lifetime $T_\mathcal{S}$. The lifetime $T_\mathcal{S}$ is given by the sum of the inter-visit times $\tilde{\tau}_k$ ($k<n$) corresponding to $\tau_k$ conditioned on being lesser than the time $\mathcal{S}$ to starve. 
With this the distribution of $T_\mathcal{S}$ reads:
\begin{align}
\label{eq:TandN}
    &\mathbb{P}(T_\mathcal{S}=t) \nonumber \\\nonumber
    &=\int_0^\infty {\rm d}n \mathbb{P} \left( \sum_{k=0}^{N_\mathcal{S}-1} \tau_k  + \mathcal{S} = t| N_\mathcal{S} = n \right) \mathbb{P}(N_\mathcal{S} = n) \\\nonumber
    & \approx \int_0^\infty {\rm d}n \delta \left( n\left\langle \tilde{\tau}_\infty \right\rangle + \mathcal{S} - t \right) \mathbb{P}(N_\mathcal{S} = n).
\end{align}
where we have used that, for large $k$, the distribution of $\tilde{\tau}_k$ becomes independent of $k$ and the sum follows the law of large numbers, $\sum\limits_{k=0}^{n-1}\left\langle \tilde{\tau}_k \right\rangle \sim n \lim\limits_{k \to \infty} \left\langle\tilde{\tau}_k \right\rangle=n  \left\langle\tilde{\tau}_\infty \right\rangle$ (see SM for details and numerical checks). This leads to the tail distribution:
\begin{align}
\label{eq:TandNgeq}
    \mathbb{P}(T_\mathcal{S} \geq t) \approx  \mathbb{P}\left(N_\mathcal{S} \geq  \frac{t-\mathcal{S}}{\left\langle\tilde{\tau}_\infty \right\rangle} \right) \; .
\end{align}
In particular, the scaling with $\mathcal{S}$ of the first two cumulants of the lifetime $T_\mathcal{S}$ is given by
\begin{align}\label{eq:momT}
    \left\langle T_\mathcal{S}\right\rangle  ,\sqrt{\text{Var}(T_\mathcal{S})} \propto
\begin{cases}
\mathcal{S} &  \mu<1\\
\mathcal{S}^2      & \mu=1\\
\exp\left[ \left( \mathcal{S}/a\right)^{\mu/(1+\mu)}\right]                       & \mu>1
\end{cases}
\end{align}
In Fig.~\ref{fig:Lifetime} we validate Eqs.~\eqref{eq:TandNgeq}-\eqref{eq:momT} via simulations. \leo{These results have strong consequences in the important case of a diffusing particle in dimension $1$, $2$ and $3$: a given metabolic time $\mathcal{S}$ leads to radically different lifetimes depending on the space dimension (see SM Fig.~S7 for comparison of the average lifetimes).} \\

We show in SM that the knowledge of the lifetime  distribution allows one to obtain the distribution of the position of the forager at starvation, denoted by $\vec{R}_\mathcal{S}$ and defined for the isotropic RWs considered here by its norm $R_\mathcal{S}$. In particular, we find that (up to log corrections in marginal case)
\begin{align}\label{eq:momR}
    \left\langle R_\mathcal{S}\right\rangle  ,\sqrt{\text{Var}(R_\mathcal{S})} \propto
\begin{cases}
\mathcal{S}^{1/d_\text{w}} &  \mu<1\\
\mathcal{S}^{2/d_\text{w}}      & \mu=1\\
\exp\left[\frac{1}{d_\text{w}}\left( \frac{\mathcal{S}}{a} \right)^{\frac{\mu}{1+\mu}} \right]                       & \mu>1.
\end{cases}
\end{align}

Several comments  on Eqs. \eqref{eq:momN}-\eqref{eq:momR}, echoing the results of the $1d$ nearest neighbour starving RW \cite{benichou2014depletion,Benichou_2016} case recalled in the introduction, are in order. (i) The fluctuations of $N_{\mathcal{S}}, \;T_{\mathcal{S}}$ and $R_{\mathcal{S}}$ are relevant for all RWs classes, as the average and the standard deviation  grow similarly. This highlights the importance of the distribution, Eq.~\eqref{dist_N_S} (and Eqs.~\eqref{eq:x}-\eqref{eq:TransDistMn}). (ii) Averages of these observables grow algebraically for  recurrent (and marginal) RWs. This generalizes the  case of a regular 1d starving RW  to the case of general recurrent RWs.  Strikingly,  the mean lifetime  is  linear with the starvation index ${\cal S}$ for all recurrent walks (independently of $\mu$).  On the other hand,  the averages of  $N_{\mathcal{S}}, \;T_{\mathcal{S}}$ and $R_{\mathcal{S}}$ become stretch-exponentially large for transient RWs. Note that the  mean-field exponential behavior is, as it should, recovered in the limit $d_\text{f}\to\infty$. (iii)  The distributions of all these observables  obey asymptotically a single-parameter scaling for any value of $\mu$. This extends the  result known for $d_\text{w}=2$ in the $1d$ case \cite{benichou2014depletion,Benichou_2016} to general Markovian starving RWs. 

We have shown that the maximum of the inter-visit times of general Markovian RWs assume simple, universal limit distributions.  We have determined the corresponding rescaled variables and the asymptotic of the limit distributions.  Beyond this fundamental aspect, we have shown that these results have applications in foraging theory. They have enabled us  to determine the statistics of a variety of key observables of the $d$ dimensional starving  RW problem, and to reveal their universal features. While the universality breaks down when the inter-visit times correlations cannot be neglected, as is the case for smooth processes (such as the Random Acceleration Process \cite{Godreche:2021}, see SM), our results hold for RWs on graphs with strongly inhomogeneous degree distributions (see SM for the typical example of the $(u,v)$ flowers \cite{Rozenfeld:2007,Meyer:2012}). \\

{\it Acknowledgments.--} We thank  J. Br\'emont, J. Klinger, P. Krapivsky and S. Redner for useful discussions. \\
\\


\begin{thebibliography}{56}
\bibitem [{fir()}]{first}%
  \BibitemOpen
  \href@noop {} {}\bibinfo {note} {The exact values are $n=20691380$,
  $127427498$ and $316227766$}\BibitemShut {NoStop}%
\bibitem [{sec()}]{second}%
  \BibitemOpen
  \href@noop {} {}\bibinfo {note} {The exact values are $n=37275$, $193069$ and
  $10^6$}\BibitemShut {NoStop}%
\end{thebibliography}
%

\end{document}


\RestyleAlgo{ruled}
\title{SUPPLEMENTARY MATERIAL\\
From Maximum of inter-visit Times to Starving Random Walks}

\author{L. R\'egnier}
\author{M. Dolgushev}
\author{O. B\'enichou}
\maketitle

\tableofcontents

\section{Distribution of the maximum inter visit time $M_n$}
In this section, we derive the asymptotic distribution of the maximum $M_n$ of the inter-visit times  based on the assumption that correlations between inter-visit times $\{\tau_k\}$ can be neglected as it is \leo{self-consistently checked} in the main text and  confirmed numerically in Sec. \ref{sec:Indep}. This assumption allows to factorize the cumulative distribution function (CDF) of $M_n$ as
\begin{align}
    \mathbb{P}(M_n\leq T)&=\mathbb{P}(\max(\tau_0,\ldots,\tau_{n-1})\leq T)\\
    &\approx \prod_{k=0}^{n-1}\mathbb{P}(\tau_k\leq T) \label{eq:IndepApprox} \\
    &=\prod_{k=0}^{n-1}\left(1-\int_T^\infty F_k(\tau) {\rm d}\tau \right),  \label{eq:MaxSurvi}
\end{align}
where $F_k(\tau)$ is the distribution of $\tau_k$. Based on the ratio between the fractal $d_\text{f}$ and walk $d_\text{w}$ dimensions, $\mu \equiv d_\text{f}/d_\text{w}$, we distinguish  between the recurrent ($\mu<1$), transient ($\mu>1$), and marginal ($\mu=1$) RW classes.

\subsection{Recurrent RWs}

In the recurrent case, the inter-visit time distribution takes a scaling form \cite{regnier2023universal}, namely
\begin{align}
    F_k(\tau)=\frac{1}{k^{1+1/\mu}}\psi(\tau/k^{1/\mu})
\end{align}
with $\psi$  a process dependent scaling function, such that 
\begin{align}
\label{eq:small_psi_rec}
    \psi(x)\propto 
    \begin{cases}
        \exp\left[- A x \right] \mbox{ for } x \gg 1 \\
        1/x^{\mu+1} \mbox{ for } x \ll 1,
    \end{cases}
\end{align}
where $A$ is a process dependent constant.
For the survival probability $S_k(T)\equiv \int_T^\infty F_k(\tau) {\rm d}\tau$, this leads to:
\begin{align}
    S_k(T)=\frac{1}{k}\Psi(T/k^{1/\mu}) \; ,
\end{align}
where $\Psi$ is another process dependent scaling function, such that 
\begin{align}
\label{eq:psi_rec}
    \Psi(x)\propto 
    \begin{cases}
        \exp\left[- A x \right] \mbox{ for } x \gg 1 \\
        1/x^{\mu} \mbox{ for } x \ll 1,
    \end{cases}
\end{align}
Consequently, the cumulative distribution of $M_n$  also adopts a scaling form in the large $T$ and $n$ limit as (see Eq.~\eqref{eq:MaxSurvi})
\begin{align}
    \mathbb{P}(M_n\leq T)&=\prod_{k=0}^{n-1}\left(1-S_k(T) \right) \\
    &\sim \exp \left[-\sum_{k=0}^{n-1} S_k(T) \right] \\
    &\sim \exp \left[-\int_0^n \frac{{\rm d} k}{k} \Psi(T/k^{1/\mu}) \right] \\
    &= \exp \left[-\int_0^{n/T^{\mu}} \frac{{\rm d} x}{x} \Psi(x^{-1/\mu}) \right] \equiv \Xi_\mu (T/n^{1/\mu}).  \label{eq:derivationMnScale}
\end{align}
Thus, all the cumulants  of $M_n$  scale as $n^{1/\mu}$.

Let us now consider the limit behaviours of $\Xi_\mu(x)$ in the limits $x \equiv T/n^{1/\mu} \ll 1$ and $ x \gg 1$.\\
First, in the limit $x \gg 1$, using Eq.~\eqref{eq:psi_rec},
\begin{align}
    -\ln \Xi_\mu (x)&\propto \int_0^{1/x^{\mu}} \frac{{\rm d}x'}{x'} \exp\left[-A/x'^{1/\mu} \right]\\
    & \propto \int_0^{x} \frac{{\rm d}x'}{x'} \exp\left[-A x' \right]=E_1(Ax)\propto e^{-Ax}/x
\end{align}
where $E_1$ is the exponential integral function $E_1(X)\equiv \int_X^\infty du e^{-u}/u$.\\ 
Then, in the limit $x \ll1$
\begin{align}
    -\ln \Xi_\mu (x)&\propto  \text{const.}+ \int_1^{1/x^{\mu}} \frac{{\rm d}x'}{x'} x' \propto  x^{-\mu}.
\end{align}
To summarize, we have the two limit behaviours of $\Xi_\mu(x)$, 
\begin{align}
    -\ln \Xi_\mu (x) \propto 
    \begin{cases}
        e^{-Ax}/x &\mbox{ for } x \gg 1 \\
        x^{-\mu} &\mbox{ for } x\ll 1 
    \end{cases}
    \label{eq:RecDistTn}
\end{align}
This result is consistent with what was found in \cite{benichou2014depletion} for the specific case of a RW with nearest-neighbour jumps on the $1d$ line.

\subsection{Transient RWs}
For  transient RWs, the distribution $F_n(\tau)$ is no longer scale-invariant, and  obeys \cite{regnier2023universal}
\begin{align}
\label{rappeltransient}
    - \ln F_n(\tau) \propto
    \begin{cases}
          \tau^{\mu/(1+\mu)} &\mbox{ if } \tau \ll n^{1+1/\mu} \\
         \tau/n^{1/\mu} &\mbox{ if } \tau \gg n^{1+1/\mu}
    \end{cases}
\end{align}
A key remark is that the exponential regime is observed only for times greater than the crossover time $\tau\sim n^{1+1/\mu}$. We now use that the maximum of $n$ independent random variables of stretched exponential tails of exponent $\delta=\mu/(1+\mu)$ has average $\left\langle M_n \right\rangle \propto (\ln n)^{1/\delta}=(\ln n)^{1/\mu+1}$ and standard deviation $\sqrt{\text{Var}(M_n)}\propto \left( \ln n \right)^{1/\delta-1}=\left( \ln n \right)^{1/\mu}$ \cite{MAJUMDAR2021}. By observing that the random variable $M_n$ has a peaked distribution around its mean (because its standard deviation is small compared to its average), and that it grows much more slowly (logarithmically) than the crossover time $n^{1+1/\mu}$, we conclude that the exponential tail of the distribution \eqref{rappeltransient} does not contribute to the distribution of the maximum $M_n$. The limit law of $M_n$ (centered on its average and rescaled by its variance) is thus a classical Gumbel distribution \cite{Gumbel:1958} of zero mean and unit variance, of cumulative that we note $\Xi_\infty$ given by
\begin{align}
    -\ln \Xi_\infty (x)=\exp\left[-\pi x/\sqrt{6}-\gamma_E \right]  
    \label{eq:TransDistTn}
\end{align}
where $x=(M_n-\langle M_n \rangle)/\sqrt{\text{Var}(M_n)}$ and $\gamma_E$ is the Euler $\gamma$-constant.

\subsection{Marginal RWs}
For the marginal case,  a more detailed characterization of the tail distribution $S_n(T)\equiv \mathbb{P}(\tau_n\geq T)$ of the inter-visit time $\tau_n$ than the one provided in \cite{regnier2023universal} is needed. In particular, one has to characterize the cross-over regime $T\sim \sqrt{n}$.

\subsubsection{Summary of the known results in the marginal case}

In \cite{regnier2023universal}, it was found that the key observable leading to the different time regimes of the inter-visit time statistics is the radius of the largest ball fully visited within the visited domain. Its distribution, noted $Q_n(r)$, was found to be given by a Poisson distribution in the volume $r^{d_\text{f}}$ (which contains now an algebraic prefactor skipped in \cite{regnier2023universal})
\begin{align}
    Q_n(r)\propto \frac{r^{d_\text{f}-1}}{\rho_n^{d_\text{f}}}\exp \left[-a \left( r/\rho_n \right)^{d_\text{f}}\right]\label{eq:Qnr}
\end{align}
where the typical correlation length between traps' location $\rho_n$ is given by
\begin{align}
    \rho_n =n^{1/2d_\text{f}},
    \label{eq:scaleLargest}
\end{align}
up to omitted log corrections (see also \cite{Dembo2007}).

\subsubsection{Radius of the $k^{\rm th}$ largest visited ball}
\begin{figure}
    \centering
    \includegraphics[width=0.3\columnwidth]{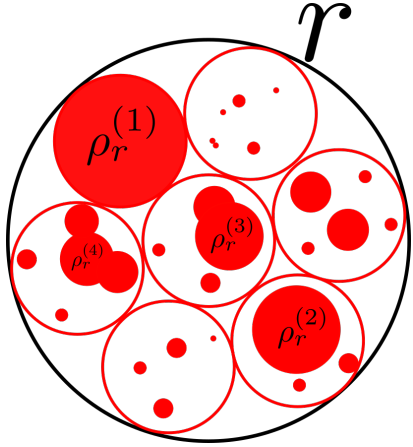}
    \caption{Illustration of the four largest fully visited balls (having radii $\rho_r^{(k)}$, numbered consecutively, $k=1,\dots,4$) by the walk before exiting the domain of radius $r$.}
    \label{fig:LargestBalls}
\end{figure}

To obtain the more detailed statistics of the random variable $\tau_n$ needed here, we search for the scaling with $k$ of the radius of the $k^\text{th}$ largest fully visited ball when $n$ sites have been visited, that we note $\rho_n^{(k)}$ (such that $\rho_n^{(1)}=\rho_n$). To do so, we will use a similar argument to the one presented in \cite{regnier2023universal}. We show that (up to log corrections in both $k$ and $n$)
\begin{align}
    \rho_n^{(k)}\propto (n/k)^{1/2 d_\text{f}} \propto \rho_n/k^{1/2 d_\text{f}}\; .
    \label{eq:Kthlargest}
\end{align}
This result is equivalent to the statement that the radius $\rho_r^{(k)}$ of the $k^{\rm th}$ largest ball entirely covered by a marginal RW before exiting the ball of radius $r$ behaves typically as $\sqrt{r/k^{1/d_\text{f}}}$, using that $r\sim n^{1/d_\text{f}}$. To obtain this result, we adapt the proof of \eqref{eq:scaleLargest} given in \cite{regnier2023universal,Dembo2007} to cover the case of the successive largest fully visited balls (see \ref{fig:LargestBalls} for an illustration).\\

We start with dividing the ball of  radius $r$ into non-overlapping balls, each of  radius $r^\gamma$, where $0 < \gamma < 1$. The objective is to determine the largest value of $\gamma$ for which at least one ball of  radius $r^\gamma/e^2$ is entirely visited by the RW before leaving the original ball of radius $r$.  Within this spherical region, there are $r^{d_\text{f}-\gamma d_\text{f}}$ disjoint spheres, each having a radius $r^\gamma$. Our initial inquiry revolves around determining how many times the RW will penetrate one of these $r^\gamma$-radius spheres before departing from the larger sphere of radius $r$. The splitting probability associated with a circle of radius $a$ before encountering a circle of radius $b$, for a RW starting from  a distance $\rho$ from the center, provides  the answer to this question. It is given by \cite{redner_2001}:
\begin{align}
    \pi (\rho ) =\frac{\ln(\rho/b)}{\ln(a/b)}  \; . \label{eq:split2D}
\end{align} 
We define a $\gamma$-incursion for a walker initiating its trajectory from the surface of a sphere with  radius $\rho = r^\gamma$ as the event where the walker enters a sphere with radius $a = r^\gamma / e$ before reaching a sphere with radius $b = r$. The probability associated with this specific type of incursion can be expressed as follows:
\begin{align*}
  \frac{\ln(r^\gamma/r)}{\ln(r^\gamma/(e\,r))}\sim 1-\frac{1}{(1-\gamma)\ln r }\;.  
\end{align*} 
For one sphere of radius $r^\gamma$, the probability of making $x$ $\gamma$-incursions is 
\begin{equation}
    \mathbb{P}(X=x)= \left( 1-\frac{1}{(1-\gamma)\ln r}\right)^{x} \sim \exp\; \left( -\frac{x}{(1-\gamma)\ln r} \right)
\end{equation}
 Thus, the distribution of the number of incursions is exponential.

We proceed by making the assumption that the number of $\gamma$-incursions for each of the $M = r^{d_\text{f}-\gamma d_\text{f}}$ individual spheres is a set of independent and identically distributed (i.i.d.) random variables, denoted as $X_m$ for $m = 1, \ldots, M$. These random variables are exponentially distributed with an average value of $\left\langle X \right\rangle = (1-\gamma)\ln r$. Consequently, the average value of the $k^{\rm th}$ maximum number of $\gamma$-incursions, denoted as $\mathcal{M}_M^{(k)}$, is expressed as follows (see \cite{renyi1953theory}):
\begin{align}
    \left\langle \mathcal {M}^{(k)}_M \right\rangle &
     \sim \left\langle X \right \rangle \ln \left( M/k \right) \; .
     \label{eq:Max_rv}
\end{align}
Using the dependence of $M$ and $\left\langle X \right \rangle$ on $r$, the $k^{\rm th}$ largest number of $\gamma$-incursions inside a ball of radius $r^\gamma$ reads: 
\begin{align}
    (1-\gamma)^2d_\text{f} \left(  \ln r \right) \left( \ln \left(r/k^{1/(1-\gamma)d_\text{f}} \right) \right)
\end{align}

We shall now investigate the number of incursions required inside a sphere with radius $r_0^\gamma$ to ensure the visitation of all the sites contained within a sphere of radius $r_0^\gamma/e^2$. We begin by calculating the probability of reaching the origin during a $\gamma$-incursion, taking into account that the incursion terminates upon the random walk intersecting the sphere with radius $r_0^\gamma$ (where another $\gamma$-incursion may potentially commence). Employing Eq.~\eqref{eq:split2D} for the splitting probability, with an outer radius $b = r_0^\gamma$, an inner radius $a = 1$, and an initial position at a radius $\rho = r_0^\gamma/e$ (which marks the commencement of the $\gamma$-incursion), we get the probability of successfully reaching the origin within this incursion to be $1/(\gamma \ln r_0)$. Consequently, the probability of not reaching the origin during any of the $x$ $\gamma$-incursions is:
\begin{align*}
 \left(1-\frac{1}{\gamma \ln r_0} \right)^x \sim \exp \left( -\frac{x}{\gamma \ln r_0}\right)\;.
\end{align*} 
As the ball with radius $r_0^\gamma/e^2$ encompasses $(r_0^\gamma /e^2)^{d_\text{f}}$ sites, we can estimate the probability of having at least one unvisited site after performing $x$ incursions as follows:
\begin{align}
&1-\mathbb{P}(\text{all sites are visited in the ball of radius $r_0^\gamma/e^2$})\nonumber\\
    &\approx 1-\mathbb{P}(\text{$0$ is visited})^{(r_0^\gamma/e^2)^{ d_\text{f}}}\nonumber\\
    &=1-\left(1-\exp\left(-\frac{x}{\gamma \ln r_0} \right)\right)^{(r_0^\gamma/e^2)^{ d_\text{f}}}\nonumber\\
    &\approx (r_0^\gamma/e^2)^{ d_\text{f}} \times \exp \left( -\frac{x}{\gamma \ln r_0}\right)=\frac{1}{e^{2d_\text{f}}}\exp \left( \gamma \,d_\text{f}\, \ln r_0-\frac{x}{\gamma \ln r_0}\right).
\end{align} 
Here, we make the assumption that all $(r_0^\gamma/e^2)^{d_\text{f}}$ sites within the sphere are equivalent to the origin, and the exploration of these particular sites is regarded as independent events. The probability of having at least one unvisited site after $x$ incursions diminishes as $x$ increases, eventually approaching zero for $x$ exceeding a critical value given by:
\[x_c(r) = \gamma^2 d_\text{f} (\ln r_0)^2.\]
Therefore, it can be inferred that a total of $\gamma^2 d_\text{f} (\ln r_0)^2$ incursions are requisite to visit all the sites contained within the sphere of radius $r_0^\gamma$.

Finally, given that there are $(1-\gamma)^2d_\text{f} \left(\ln r \right) \left( \ln \left(r/k^{1/(1-\gamma)d_\text{f}} \right) \right)$ incursions within the $k^{\rm th}$ maximally visited ball and that $\gamma^2\,d_\text{f}\,(\ln r_0)^2$ incursions are required to completely explore the ball with radius $r_0^\gamma/e^2$ nested within this $k^{\rm th}$ maximally visited ball with radius $r^\gamma$, it follows that the radius $\rho_r^{(k)}=r_0^\gamma$ must satisfy the condition:
\begin{align}
    (1-\gamma)^2d_\text{f} \left(  \ln r \right) \left( \ln \left(r/k^{1/(1-\gamma)d_\text{f}} \right) \right)=\gamma^2\,d_\text{f}\,(\ln r_0)^2
\end{align}
In the limit $\ln k \ll \ln r$ ($k$ grows more slowly than any power of the radius), taking $r_0=r/k^a$ (with $a$ unknown) this implies 
\begin{align}
    \gamma=1/2 , \; a=1/d_\text{f}
\end{align}
and we find the announced Eq.~\eqref{eq:Kthlargest} by taking $r=n^{1/d_\text{f}}$.

\subsubsection{Statistics of $\tau_n$ in the marginal case}
In this section, we focus on the determination of the survival probability $S_n(T)$ for $T\sim \sqrt{n}$. Indeed, we expect the statistics in this time scale to dominate \cite{regnier2023universal} as the tail distribution of the inter-visit times is algebraic at times smaller than $\sqrt{n}$ (which makes the maximum of the inter-visit times large \cite{Bouchaud:1990}) and stretched exponential after (which imposes a cut-off for the maximum statistics, see also Eqs.~\eqref{eq:RecDistTn} and \eqref{eq:TransDistTn}).\\
In order to obtain an accurate estimation of the tail distribution of $\tau_n$, we start by dividing the visited territory in subregions (balls) ordered consecutively by their size ($k=1$ corresponds to the largest fully visited ball, $k=2$ the next one, etc, see \ref{fig:LargestBalls}). We define $S_n^{(k)}(T)$ as the probability to start in the $k^{\rm th}$ largest fully visited ball and not to visit any new site up to time $T$. For $T\sim \sqrt{n}$, we partition the survival event over the $k^{\text{th}}$ largest fully visited subregion it starts from (of note, there are at most $n$ disjoint visited balls): 
\begin{align}
    S_n(T)=\sum_{k=1}^{n} S_n^{(k)}(T)
    \label{eq:SplitMarg}
\end{align}
Then, $S_n^{(k)}(T)$ can be partitioned over the radius of the $k^{\rm th}$ maximally visited ball,
\begin{align}
    S_n^{(k)}(T)=\int_0^{n^{1/d_\text{f}}}{\rm d}r  P_n(r)Q_n^{(k)}(r) S_r(T) \frac{1}{T},
    \label{eq:SurvBalls}
\end{align}
where 
\begin{itemize}
     \item $P_n(r)$ is the probability to start the visitation process of the $n^{\rm th}$ site from a region of radius $r$ containing visited sites only. Because the surface of the visited domain is $\propto n$ and the number of sites within the circle of radius $r$ is $r^{d_\text{f}} \ll n$ (in the marginal case, the volume of the maximal sphere with only visited sites is $\propto \sqrt{n}$), we have that $P_n(r)\propto \frac{r^{d_\text{f}}}{n}$ by assuming the RW starts uniformly inside the visited domain (as assumed in the proof \cite{Dembo2007}).
    \item $Q_n^{(k)}(r)$ is the probability to have a typical radius $r$ for the $k^{\rm th}$ maximal visited region. It is a scaling function of $x=r/\rho_n^{(k)}$, $Q_n^{(k)}(r)=\frac{r^{d_\text{f}-1}}{(\rho_n^{(k)})^{d_\text{f}}}f(r/\rho_n^{(k)})$. In particular, by interpreting $\rho_n^{(k)}$ as a correlation length between location of traps, one has, in line with Eq.~\eqref{eq:Qnr}, that $f(x)$ decays exponentially with $x^{d_\text{f}}$ (exponential decay in the rescaled volume).
    \item $S_r(T)$ is the probability not to have exited the $k^{\rm th}$ maximal visited region of radius $r$ up to time $T$, which is $\exp \left[-b T/r^d \right]$ (by noting $d=d_\text{w}=d_\text{f}$) at large times ($b$ is a model dependent constant).
    \item $1/T$ is the probability not to have visited any new site up to time $T$ conditioned on having stayed inside the $k^{\rm th}$ maximal visited region, in agreement with the result found in \cite{regnier2023universal} (the visited region is fractal and semi-infinite due to the conditioning).
\end{itemize}

In ~\ref{fig:Calc2d} \textbf{(a)} and \textbf{(b)}, we check the functional forms of $P_n(r)$ and $Q_n^{(k)}(r)$ above, respectively, for the marginal L\'evy flight in $1d$ ($\alpha=1$) as well as in \textbf{(c)} the average area of the $k^\text{th}$ largest fully visited domains for the $2d$ nearest-neighbour RW. 
Overall, ~\ref{fig:Calc2d} validates the main ingredients of Eq.~\eqref{eq:SurvBalls}, where we made the further assumptions of independence between the exact position in the ball and time to find a new site inside or outside the ball. 
\begin{figure}
    \centering
    \includegraphics[width=\columnwidth]{Calc_marg.pdf}
    \caption{{\bf Numerical checks of the key hypotheses involved in Eq.~\eqref{eq:SurvBalls}:}\\  {\bf (a)} For a $1d$ L\'evy flight of parameter $\alpha=1$, probability to start in the unvisited interval of maximal radius $r$, $P_n(r)$, times the number of visited sites $n$  (blue, orange, and green symbols represent $n=400$, $600$, and $800$, respectively, the dashed line being a linear function of $r$).\\ {\bf (b)} For a $2d$ nearest neighbour RW, tail probability distribution of having a fully visited domain of $k^\text{th}$ largest area $A$, $S_n^{(k)}(A)=\int_{r^2\geq A}Q_n^{(k)}(r) d^{d_\text{f}}r$  ($k=1$, $2$, and $4$ in blue stars, orange circles, and green squares, respectively). We use the rescaled variable $\sqrt{k}A/(\rho_n^{(1)})^2$, where $\rho_n^{(1)}\equiv\rho_n$ is the average length of the largest unvisited interval, and $n=10^6$.\\ {\bf (c)} For a $2d$ nearest neighbour RW, the average area of the $k^\text{th}$ largest fully visited square as a function of $x=k/n$ (dashed line corresponds to $1/\sqrt{x}$), for $n=10^4$, $10^5$, and $10^6$ (blue stars, orange circles, and green squares, respectively). }
    \label{fig:Calc2d}
\end{figure}

For times $T$ smaller than the typical exit time $\rho_n^{d}=\sqrt{n}$ from the largest region maximally visited, $T \lesssim \sqrt{n}$, we have two types of terms in Eq.~\eqref{eq:SplitMarg}: Using Eq.~\eqref{eq:SurvBalls}, one realizes that either the time $\left( \rho_n^{(k)} \right)^d$ to exit the $k^\text{th}$ largest sphere is larger than $T$  (so we are still in the limit of a semi-infinite medium) or it is smaller (the walker almost surely exits the domains). In the first case, we have $S_n^{(k)}(T)\approx \int_0^\infty  P_n(r) Q_n^{(k)} {\rm d} r \frac{1}{T}\approx \frac{1}{\sqrt{n k}T} $, and in the second case, it vanishes (due to the exponential cut-off). This implies that 
\begin{align}
    S_n(T)&\propto \sum_{(\rho_n^{(k)})^d=\sqrt{n/k}\geq T} \frac{1}{\sqrt{n k}T} \\
    &\propto \frac{1}{\sqrt{n}T}\sqrt{\frac{n}{T^2}}=\frac{1}{T^2} 
    \label{eq:Marg_short}
\end{align}

For $T \gtrsim \sqrt{n} $ (up to log corrections), as we proceed to show, the survival probability of the walks starting in the largest ball dominates.
Considering the first term $S_n^{(1)}(T)$, we resort to the Laplace method using that $S_r(T)\propto \exp\left[- b T/r^{d} \right]$ and $Q_n(r) \propto \frac{r^{d-1}}{\rho_n^d}\exp \left[-a r^d/\rho_n^d \right]$ which results in 
\begin{align}
    S_n^{(1)}(T)&\int_0^{n^{1/d}}{\rm d} r Q_n(r) P_n(r) \frac{1}{T} S_r(T) \\
    &\propto \int_0^{n^{1/d}} \frac{r^{d-1}{\rm d} r}{\rho_n^d}  \frac{r^d}{n} \frac{1}{T}\exp\left[-ar^d/\rho_n^d-bT/r^d \right] \\
    &=\frac{1}{n}\int_0^{n^{3/4d}/T^{1/2d}} \rho^{2d-1} {\rm d} \rho \exp\left[-\sqrt{T/\rho_n^d}\left(a\rho^d+b/\rho^d \right) \right] \\
    &\propto \frac{1}{n} f(T/\rho_n^d)\exp\left[-\sqrt{\frac{T}{\rho_n^d}} U(\rho^*) \right] \; , \label{eq:Marg_long}
\end{align}
$U(\rho)=a \rho^d +b/\rho^d$, $U(\rho^*)$ stands for its minimum, and $f(x)=1/x^{1/4}$. In fact, the other terms in Eq.~\eqref{eq:SplitMarg} can be neglected. Indeed, for the $k^\text{th}$ term $S_n^{(k)}(T)$ ($k>1$), we obtain via the same method that
\begin{align}
    \int_0^{n^{1/d}}{\rm d} r Q_n^{(k)}(r)\frac{1}{T} S_r(T) P_n(r) \sim \frac{B}{n} f(T/\rho_n^d)\exp\left[-\sqrt{\frac{T}{\rho_n^d}} U^{(k)}(\rho^{*(k)}) \right]
\end{align}
where $U^{(k)}(\rho)=\sqrt{k}a \rho^d +b/\rho^d$ and $U^{(k)}(\rho^{*(k)})=k^{1/4}U(\rho^*)$ its minimum. In particular, this implies that
\begin{align}
    \frac{\int_0^{n^{1/d}}{\rm d} r Q_n^{(k)}(r) \frac{1}{T}S_r(T) P_n(r)}{ \int_0^{n^{1/d_\text{f}}}{\rm d} r Q_n(r) P_n(r) S_r(T)} \propto \exp \left[-(k^{1/4}-1)U(\rho^*)\sqrt{T/\rho_n^d} \right] \ll 1
\end{align}
Thus, for $T\gtrsim \sqrt{n}$, one can neglect  the fully visited subregions which do not have the maximal size. \\

Finally, combining Eqs.~\eqref{eq:Marg_short} and \eqref{eq:Marg_long}, we get that for $T\sim \sqrt{n}$,
\begin{align}
    S_n(T)&=\frac{1}{n}\Psi_1(T/\sqrt{n}) .\label{eq:2dScaleInv}
\end{align}
This functional form will be used for derivation the distribution of the maximum. It has the following asymptotic behaviors:
\begin{align}
     S_n(T) \propto \frac{1}{n}\Psi_1(T/\sqrt{n}) \propto 
    \begin{cases}
          \frac{1}{T^2} &\mbox{ if } \sqrt{n} \lesssim T \\
          \frac{1}{n} \left(\frac{\sqrt{n}}{T}\right)^{1/4} \exp\left[-\text{const.} \sqrt{T/\sqrt{n}} \right]    &\mbox{ if } \sqrt{n} \gtrsim T
    \end{cases}\label{eq:Sbeh}
\end{align}

\subsubsection{Distribution of the maximum for the marginal RW}

Based on the functional form \eqref{eq:Sbeh} for $S_n(T)$, $M_n/\sqrt{n}$ converges to a non-trivial distribution at large $n$. For $x$ real and positive,
\begin{align}
    \mathbb{P}(M_n \leq x \sqrt{n})\approx \exp \left[ - \int_0^n  S_k(x \sqrt{n}) {\rm d } k \right] .
\end{align}
Then, for $x=O(1)$ (for simplicity, we put all the constant prefactors separating the different regimes of $S_n(T)$ at $1$), 
\begin{align}
    \mathbb{P}(M_n \leq x \sqrt{n})&\approx \exp \left[ - \int_0^{x^{2/3}n^{1/3}} \exp\left[- \text{const. } x\sqrt{n}/k \right] {\rm d } k - \int_{x^{2/3}n^{1/3}}^n \Psi_1\left(\sqrt{\frac{nx^2}{k}}\right) \frac{ {\rm d } k}{k} \right]  \\
    &\approx  \exp \left[ - \int_0^{1/x^2} \Psi_1\left(\frac{1}{\sqrt{u}}\right) \frac{ {\rm d } u}{u} \right] \; . \label{eq:distMmarg}
\end{align}
From this, we deduce that $M_n/\sqrt{n}$ has a limit distribution. We note that $M_n$ does not have a scale invariant form as was the case of recurrent RWs, however, it still has asymptotically a single parameter scaling as the distribution of $M_n/\left\langle M_n \right\rangle $ converges to a cumulative distribution we note $\Xi_1$ ($\left\langle M_n \right\rangle \sim \sqrt{\text{Var}(M_n)}\sim \sqrt{n}$ up to log prefactors). Based on Eqs.~\eqref{eq:Sbeh} and \eqref{eq:distMmarg}, we  obtain (up to log corrections)
\begin{align}
    -\ln \Xi_1 (x) \propto 
    \begin{cases}
        1/x^{2} &\mbox{ for } x \ll 1 \\
        e^{-Bx^{1/2}} &\mbox{ for } x\gg 1 
    \end{cases},
    \label{eq:MargDistTn}
\end{align}
where $B$ being a process dependent constant.
\subsection{Additional examples}
\begin{figure}
    \centering
    \includegraphics[width=\columnwidth]{DistribMn_additional.pdf}
    \caption{{\bf Additional examples of $M_n$ distributions:}\\  
    {\bf (a)}-{\bf (c)} Distributions of the maximum
     $M_n$ as a function of the rescaled variable $x$, defined in Eq. (1) of the main text for $1d$ L\'evy flights of parameter {\bf (a)} $\alpha=1.5$ (recurrent) {\bf (b)} $\alpha=1$  (marginal) {\bf (c)} $\alpha=0.5$ (transient, here $m_n=\frac{M_n-\left\langle M_n \right\rangle }{\sqrt{\text{Var}(M_n)}}$). Increasing values of $n$ are represented by blue circles, orange stars and green squares:  {\bf (a)}  $n= 1389$, $3727$ and $10^4$, {\bf (b)} $n= 545559$, $2335721$, $10^7$, {\bf (c)} $n= 48329$, $263665$, $1438449$.  In {\bf (a)} and {\bf (b)} the insets represent the distributions for small $x$ values in log-log scales.\\  {\bf (a$^\prime$)}-{\bf (c$^\prime$)} Average (blue circles) and standard deviation (orange squares) of $M_n$ for the RWs of panels {\bf (a)}-{\bf (c)}.\\ In all plots, the black dashed lines correspond to the best fit of the proportionality prefactors and process-dependent constants of the asymptotic expression of Eqs.~\eqref{eq:RecDistTn}, \eqref{eq:MargDistTn}, \eqref{eq:TransDistTn} and Eqs. (2) and (3) of the main text.  }
    \label{fig:Mn_distrib}
\end{figure}
In addition to the numerical checks of Fig. 2 of the main text, we display in ~\ref{fig:Mn_distrib} the asymptotic distribution of the maximum for $1d$ L\'evy flights (jump distribution being $p(\ell)\propto \ell^{-1-\alpha}$) of parameter $\alpha=d_\text{w}=1.5$ (recurrent), $1$ (marginal), and $0.5$ (transient). It is presented for the rescaled variable  $M_n/\left\langle M_n\right\rangle $ for $\mu\leq 1$ and $(M_n-\left\langle M_n\right\rangle)/\sqrt{\text{Var}\left(M_n \right)}$ for $\mu>1$, for different values of $n$. In addition to the figures presented in the main text, these additional examples further confirm the validity of our results. 

\section{Independence of inter visit times $\tau_k$}\label{sec:Indep}

In this section, in  addition to the argument presented in the main text, we check numerically the effective independence of the inter-visit times $\{\tau_k\}$. For this, we compare the rescaled distributions of $M_n$ and that of the maximum $M'_n$ of random variables distributed as the $\tau_k$ but which are independent, denoted by $\tau'_k$. To sample $M'_n$,  we use the following method:
\begin{itemize}
    \item First, we draw $N$ (here we take $N=100$) RW trajectories leading to $n$ visited sites and keep only the inter-visit times. We thus have $N=100$ lists $[\tau_0,\ldots,\tau_{n-1}]$ drawn independently.
    \item Then, for each $k$ from $0$ to $n-1$, we randomly select one of the $N$ lists (uniformly) and take the $k^\text{th}$ element, $\tau_k$. It results in a list $[\tau'_0,\ldots, \tau'_{n-1}]$.
    \item Finally, we have one realization of $M_n'$ as the maximum of the list $[\tau'_0,\ldots, \tau'_{n-1}]$.
\end{itemize}

\begin{figure}[th!]
    \centering
    \includegraphics[width=\columnwidth]{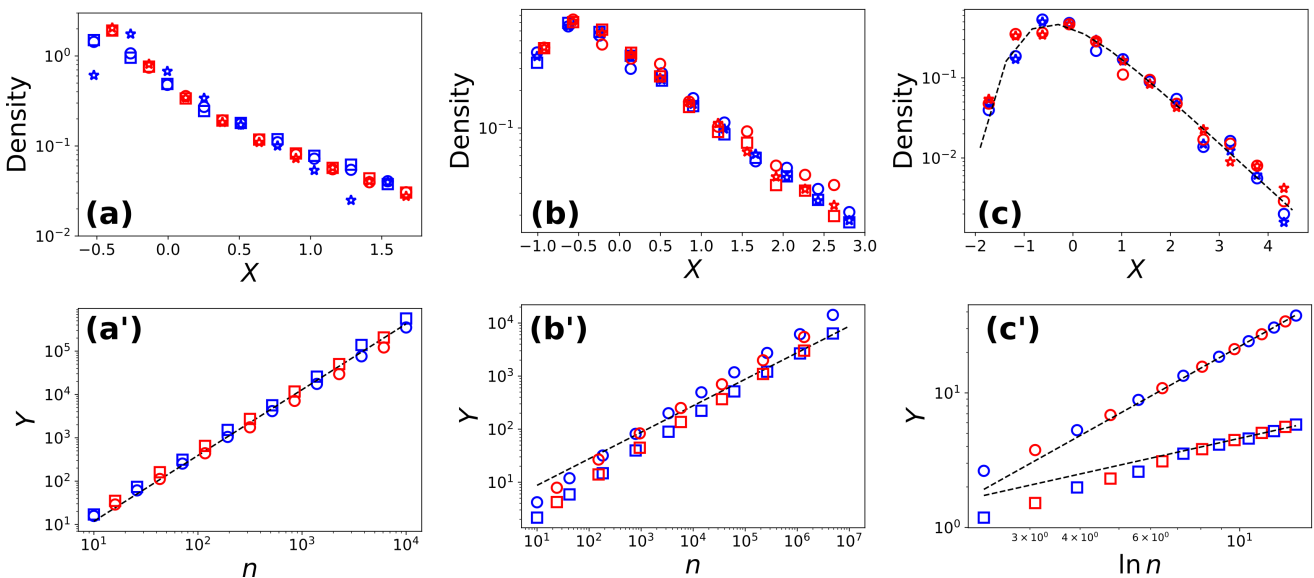}
    \caption{{\bf Numerical check of the effective independence of $\{\tau_k\}$ for their maximum.} \\
     Comparison of {\bf (a)}-{\bf (c)} the centered and normalized distributions of $M_n=\max(\tau_0,\ldots,\tau_{n-1})$ and $M'_n=\max(\tau'_0,\ldots,\tau'_{n-1})$ (the simulations with independent inter-visit times, of $M'_n$, are in blue, the dependent ones, of $M_n$, in red) as well as {\bf (a$^\prime$)}-{\bf (c$^\prime$)} their average (circles) and standard deviation (squares) for a {\bf (a)} RW on a percolation cluster, $\mu\approx 0.659$ (recurrent); $n= 2275$, $3727$, and $6105$, {\bf (b)} nearest neighbour RW in $2d$, $\mu=1$ (marginal); $n=88$, $6951$, and $545559$, {\bf (c)}  nearest neighbour RW in $3d$, $\mu=3/2$ (transient); the distribution is compared to the Gumbel of zero mean and unit variance represented by a black dashed line, $n=84384$ and $193069$. For {\bf (a)}-{\bf (c)}, increasing values of $n$ are represented by circles, stars and squares successively. The average and standard deviation are compared to their expected scaling with $n$ shown by black dashed lines.}
    \label{fig:Independence}
\end{figure}

 In line with the analytical argument given in the main text, the rescaled distributions of $M_n$ and $M'_n$ have the same behavior, as observed in the first line of \ref{fig:Independence}. Besides, the cumulants shown in the second line of \ref{fig:Independence} have the same scaling with $n$ in both cases, further confirming the relevance of the independence approximation. 

We conclude that the numerical tests validate, additionally to the analytical argument of the main text, the independence approximation made in Eq.~\eqref{eq:IndepApprox}.

\section{Distribution of the number of sites visited at starvation $N_\mathcal{S}$}
\subsection{Analytical derivation}
\label{sec:NSDerivation}

The number $N_\mathcal{S}$ of sites  visited at starvation is larger or equal to $n$ if the $n$ first inter-visit times are all smaller than $\mathcal{S}$,  ${\tau_0<\mathcal{S}},\dots,{\tau_{n-1}<\mathcal{S}}$. Thus, it means that the maximum inter-visit time is smaller than $\mathcal{S}$, and we have the following relation between $N_\mathcal{S}$ and $M_n$: 
\begin{align}
    \mathbb{P}(N_\mathcal{S}\geq n)=\mathbb{P}(M_n < \mathcal{S} ). \label{eq:inversion_M_N}
\end{align}
It implies that the knowledge of the distribution of $M_n$ 
 (whose asymptotic behavior is given in \eqref{eq:RecDistTn}, \eqref{eq:TransDistTn} and \eqref{eq:MargDistTn}) allows one to obtain the distribution of $N_\mathcal{S}$. In particular:
\begin{itemize}
    \item For recurrent RWs: The distribution of $N_\mathcal{S}$ is scale-invariant (so that $N_\mathcal{S}/\mathcal{S}^\mu$ is independent of $\mathcal{S}$) and its asymptotics is given by:
    \begin{align}
    -\ln\mathbb{P}(N_\mathcal{S}/\mathcal{S}^\mu \geq x)=-\ln \Xi_\mu (x^{-1/\mu}) \propto 
    \begin{cases}
        E_1(Ax^{-1/\mu})\propto e^{-A/x^{1/\mu}}x^{1/\mu} &\mbox{ for } x \ll 1 \\
        x &\mbox{ for } x\gg 1.
    \end{cases}
    \label{eq:NSRec}
\end{align}
    \item For transient RWs: By noting $\left\langle M_n\right\rangle \sim a (\ln n)^{\frac{1+\mu}{\mu}}$ ($a$ is a model dependent constant) and $\sqrt{\text{Var} (M_n) } \sim b(\ln n)^{\frac{1}{\mu}}\equiv\frac{a(1+\mu)\sqrt{6}}{\pi \mu} (\ln n)^{\frac{1}{\mu}}$  \cite{MAJUMDAR2021},  
    $\mathbb{P} \left(\frac{M_n-a (\ln n)^{1/\mu+1}}{b (\ln n)^{1/\mu}} \leq x \right)$
    converges to a Gumbel distribution $\Xi_\infty(x)$ of zero mean and unit variance. Then, $\mathbb{P}(N_{\mathcal{S}=x b (\ln n)^{1/\mu}+a (\ln n)^{1/\mu+1}} \geq n)$ also converges to $\Xi_\infty(x)$ at large $n$. By inverting $\mathcal{S}=x b (\ln n)^{1/\mu}+a (\ln n)^{\frac{1+\mu}{\mu}}$ in the limit of large $n$, we finally get
    \begin{align}
    \label{eq:NSTrans}
        \mathbb{P}(N_{\mathcal{S}}/\exp\left[ (\mathcal{S}/a)^{\mu/(1+\mu)}\right] \geq x) \sim \Xi_\infty\left(-\frac{\sqrt{6}}{\pi} \ln x\right)=\exp\left[ -x e^{-\gamma_E} \right].
    \end{align}
    In particular, we deduce that $N_\mathcal{S}$ obeys asymptotically a single parameter scaling, and thus its average and standard deviation scale with $\mathcal{S}$ in the same way (i.e., their ratio is finite). Furthermore, the asymptotic distribution is exponential in the number of sites visited at starvation $N_{\mathcal{S}}$.
    \item For marginal RWs: As $\mathbb{P}(N_{\mathcal{S}=\sqrt{n}x}  \geq n)=\mathbb{P}(M_n \leq \sqrt{n}x) \to \Xi_1(x) $, with $\Xi_1(x)$  of Eq.~\eqref{eq:2dScaleInv}, we get that ${\mathbb{P}(N_\mathcal{S}/\mathcal{S}^2 \geq x) \to \Xi_1(1/\sqrt{x})}$. In particular,  the distribution has a single-parameter scaling at large starvation index $\mathcal{S}$ and number of visited sites at starvation $N_\mathcal{S}$, and its average and standard deviation scale in the same way with $\mathcal{S}$. Based on Eq.~\eqref{eq:MargDistTn}, up to logarithmic corrections,
    \begin{align}
    \label{eq:NSMarg}
      -\ln \mathbb{P}(N_\mathcal{S}/\mathcal{S}^2 \geq x) \propto
      \begin{cases}
          e^{-A/x^{1/4}} &\mbox{ for } x\ll 1 \\
          x &\mbox{ for } x\gg 1.
      \end{cases}
    \end{align}
\end{itemize}
Thus, we have obtained the asymptotic distribution of the number of visited sites at starvation $N_\mathcal{S}$ for all types  of Markovian RWs. This distribution obeys a single parameter scaling in $x\equiv N_\mathcal{S}/ \left\langle N_{\mathcal{S}}\right\rangle$ whatever the value of $\mu$ (in contrast to the maximum $M_n$), where
\begin{align}
\label{eq:NSmom}
    \left\langle N_{\mathcal{S}} \right\rangle \propto 
    \begin{cases}
        \mathcal{S}^{\mu} & \mbox{ for } \mu <1 \; , \\
        \mathcal{S}^2 & \mbox{ for } \mu =1 \; , \\
        \exp\left[ \left( \mathcal{S}/a \right)^{\mu/(1+\mu)}\right] & \mbox{ for } \mu>1.
    \end{cases}
\end{align}
Consequently, $\sqrt{\text{Var} (N_{\mathcal{S}}) }$ has the same behavior as $ \left\langle N_{\mathcal{S}} \right\rangle$.

\begin{figure}[th!]
    \centering
    \includegraphics[width=\columnwidth]{DistribN.pdf}
    \caption{{\bf Distribution of the number of sites visited at starvation $N_\mathcal{S}$.}\\
    {\bf (a)}-{\bf (c)} Distributions of the number of visited sites at starvation $N_\mathcal{S}$ as a function of the rescaled variable $x=N_\mathcal{S}/\left\langle N_\mathcal{S}\right\rangle$, with $\left\langle N_\mathcal{S}\right\rangle$ defined in Eq.~\eqref{eq:NSmom} for {\bf (a)} RWs on $2d$ percolation clusters ($\mu\approx 0.659$, recurrent), {\bf (b)} $2d$ nearest neighbour RWs ($\mu=1$, marginal),  {\bf (c)} $3d$ nearest neighbour RWs ($\mu=3/2$, transient). Increasing values of $\mathcal{S}$ are represented by blue circles, orange stars, and green squares, respectively: {\bf (a)} $\mathcal{S}= 14667$, $31622$, and $6819$, {\bf (b)} $\mathcal{S}=2335$, $4832$, and $10^4$, {\bf (c)} $\mathcal{S}= 12$, $17$, and $22$. In {\bf (a)} and {\bf (b)} the insets represent the distributions for small $x$ values.\\
    {\bf (a$^\prime$)}-{\bf (c$^\prime$)} Average (blue circles) and standard deviation (orange squares) of $N_\mathcal{S}$  for the RWs of panels {\bf (a)}-{\bf (c)}.\\ In all plots, the black dashed lines correspond to the best fit of the proportionality prefactors and process-dependent constants of the asymptotic expression of Eqs.~\eqref{eq:NSRec}, \eqref{eq:NSTrans}, \eqref{eq:NSMarg}, and \eqref{eq:NSmom}.  }
    \label{fig:Explored}
\end{figure}

\subsection{Numerical check}\label{sec:NSnum}
In \ref{fig:Explored}, we check  the expressions \eqref{eq:NSRec}, \eqref{eq:NSTrans} and \eqref{eq:NSMarg} by fitting the  proportionality factors and the model dependent constants to the numerical data.
The recurrent RWs were run on critical percolation clusters. Each cluster was created using a periodic square lattice measuring 1000 by 1000 units. In this process, half of the lattice bonds were randomly removed, and subsequently, the largest cluster was identified. Simulations were executed on 31 clusters, selected based on their size which is around the median size among a total ensemble of 1000 clusters.
The starting site for each simulation was chosen randomly. The marginal and transient RWs were performed on square and cubic lattices, respectively. \ref{fig:Explored} validates the expressions for both the averages and the distributions of the number of visited sites at starvation $N_\mathcal{S}$. 


\section{Distribution of the lifetime $T_\mathcal{S}$}\label{sec:lifetime}

To obtain the distribution of the lifetime $T_\mathcal{S}=\tau_0+\ldots+\tau_{n-1}+\mathcal{S}$ (i.e. the number of steps before the RW starves), we use the following identity:
\begin{align}
    \mathbb{P}(T_\mathcal{S} \geq t)=\sum_{n=1}^{\infty} \mathbb{P}(N_\mathcal{S}=n)\mathbb{P}(\tau_0+\ldots+\tau_{n-1}+\mathcal{S} \geq t|N_\mathcal{S}=n) \; . \label{eq:TSfirststep}
\end{align}
Here, in the sum, the distribution of the lifetime is conditioned on $N_\mathcal{S}=n$. This results in effective random variables $\{\tilde{\tau}_k\}$ corresponding to the variables $\{\tau_k\}$ conditioned on being less than the starvation index $\mathcal{S}$. Then, using continuous notations for convenience, we can rewrite Eq.~\eqref{eq:TSfirststep} as
\begin{align}
    \mathbb{P}(T_\mathcal{S} \geq t)=\int_{1}^{\infty} {\rm d}n \mathbb{P}(N_\mathcal{S}=n)\mathbb{P}(\tilde{\tau}_0+\ldots+\tilde{\tau}_{n-1}+\mathcal{S} \geq t)
    \label{eq:TSsecondstep}
\end{align}

First, we note that the typical value of $N_\mathcal{S}$ (determined by its average) is large when $\mathcal{S} \gg 1$ (see Eq.~\eqref{eq:NSmom}). The distribution $\mathbb{P}(N_\mathcal{S}=n)$ is sharp around $n=\langle N_\mathcal{S}\rangle$ (see Eqs.~\eqref{eq:NSRec}, \eqref{eq:NSMarg} and \eqref{eq:NSTrans}). Thus, small $n$ values barely contribute to the integral of Eq.~\eqref{eq:TSsecondstep} as $\mathcal{S}$ tends to infinity. 

Second, we use that the distribution $F_k(\tau)$ of $\tau_k$ is independent of $k$ for $\tau<t_k$ (see \cite{regnier2023universal}), where $t_k$ is given by
\begin{align}
\label{eq:t_k}
   t_k=
    \begin{cases}
       k^{1/\mu} & \mbox{ for } \mu <1 \; , \\
        k^{1/2} & \mbox{ for } \mu =1 \; , \\
        k^{1/\mu+1} & \mbox{ for } \mu>1.
    \end{cases}
\end{align}
Using Eqs.~\eqref{eq:NSmom} and \eqref{eq:t_k}, one can check that the typical $\tilde{\tau}_k$ in \eqref{eq:TSsecondstep} have an index $k$ such that $t_k\geq\mathcal{S}$. For these high $k$-values, the distribution of $\tilde{\tau}_k$ (the same as that of $\tau_k$ but restricted to values smaller than $\mathcal{S}$) converges to that of a random variable noted $\tilde{\tau}_\infty$ (independent of $k$). 

Finally, based on the two previous points (large $n$ contribution and identical distribution for $\tilde{\tau}_k$) as well as the effective independence hypothesis of the inter-visit times, we obtain that the sum $\tilde{\tau}_0+\ldots+\tilde{\tau}_{n-1}$ follows asymptotically a law of large numbers and thus, with $H$ the Heaviside function, Eq.~\eqref{eq:TSsecondstep} becomes:
\begin{align}
    \mathbb{P}(T_\mathcal{S} \geq t)
    &\approx \int_1^\infty {\rm d}n \mathbb{P}(N_\mathcal{S}=n)H \left( n \left\langle \tilde{\tau}_\infty \right\rangle +\mathcal{S} - t \right) \label{eq:Life2} \\
    & \approx \mathbb{P}\left(N_\mathcal{S} \geq  \frac{t-\mathcal{S}}{\left\langle\tilde{\tau}_\infty \right\rangle} \right)
\end{align}

 We check the law of large number in \ref{fig:conditioned} on the RW models of Sec.~\ref{sec:NSnum} by computing the standard deviation of $\sum_{k=1}^n \tilde{\tau}_k$ divided by its average and observing the expected decay to zero associated to the law of large number. 

\begin{figure}[th!]
    \centering
    \includegraphics[width=\columnwidth]{tau_infS.pdf}
    \caption{{\bf Check of the law of large number.}\\ Standard deviation of $\sum_{k=1}^n \tilde{\tau}_k$ over its average ($\tilde{\tau}_k$ denotes the inter-visit times conditioned on being lesser than $\mathcal{S}$)  for $\mathcal{S}=10$, $100$, and $1000$ in blue circles, orange stars, and green squares, respectively. {\bf (a)}  RWs on a percolation cluster ($\mu\approx 0.659$, recurrent), {\bf (b)} $2d$ nearest neighbour RWs ($\mu=1$, marginal), {\bf (c)} $3d$ nearest neighbour RWs ($\mu=3/2$, transient).  }
    \label{fig:conditioned}
\end{figure}

We are left with the characterisation of $\tilde{\tau}_\infty$ which differs for each type of RWs.
\begin{itemize}
    \item For recurrent RWs: Using that the distribution $F_k(\tau)$ of the inter-visit time $\tau_k$ at early times behaves as ${F_k(\tau)\propto\tau^{-1-\mu}}$ (see Ref.~\cite{regnier2023universal}), we obtain that
    \begin{align}
        \left\langle \tilde{\tau}_\infty \right\rangle \propto \frac{\sum_{\tau=1}^{\mathcal{S}}\tau^{-\mu}}{\sum_{\tau=1}^{\mathcal{S}} \tau^{-1-\mu}} \sim b \mathcal{S}^{1-\mu}
    \end{align}
 where $b$ is a model dependent constant. Since $0<\mu<1$, the behavior $\mathcal{S}^{1-\mu}$ comes from the numerator, whereas the denominator leads to a constant, the distribution of $\tau_k$ being $F_k$ conditioned by $\tau_k \leq \mathcal{S}$. Consequently, by rescaling $T_\mathcal{S}$ by $\mathcal{S}$ and using Eq.~\eqref{eq:NSRec}, we have that
    \begin{align}
        \mathbb{P}(T_\mathcal{S}/\mathcal{S} \geq t)=\mathbb{P}\left(N_\mathcal{S} \geq b (t-1) \mathcal{S}^\mu \right)=\Xi_\mu((b(t-1))^{-1/\mu})
        \label{eq:RecLifetimeDist}
    \end{align}
    where $\Xi_\mu$ is defined in Eq.~\eqref{eq:RecDistTn}. We conclude that $T_\mathcal{S}$ scales with $\mathcal{S}$, whatever the value of the exponent $\mu<1$. 
    \item For transient RWs: In this case,
    $\tau_k$ has always finite moments (the inter-visit times' distribution is stretch-exponentially decaying) \cite{regnier2023universal}, such that the asymptotic value of the conditional inter-visit time at large $\mathcal{S}$ is the same as the unconditioned one, 
    \begin{align}
        \left\langle \tilde{\tau}_\infty \right\rangle  \to \left\langle \tau_\infty \right\rangle= b.
    \end{align}
    By rescaling $T_\mathcal{S}$ by $\exp\left[ (\mathcal{S}/a)^{\mu/(1+\mu)}\right]$, we get that
    \begin{align}
    \label{eq:TransLifetimeDist}
        \mathbb{P}(T_\mathcal{S}/\exp\left[ (\mathcal{S}/a)^{\mu/(1+\mu)}\right] \geq t)=\mathbb{P}\left(N_\mathcal{S} \geq t \exp\left[ (\mathcal{S}/a)^{\mu/(1+\mu)}\right] / b \right)=\Xi_\infty\left(-\frac{\sqrt{6}}{\pi} \ln (t/b)\right)=\exp\left[-\frac{t}{b}e^{-\gamma_E} \right]
    \end{align}
    where $a$ is defined in Sec.~\ref{sec:NSDerivation}.  The lifetime $T_{\mathcal{S}}$  is  exponentially distributed (as  the number $N_{\mathcal{S}}$ of sites visited at starvation).
    \item For marginal RWs: Similarly to  recurrent RWs, we get the average value of $\tilde{\tau}_\infty$ as
    \begin{align}
        \left\langle \tilde{\tau}_\infty \right\rangle \propto \frac{\sum_{\tau=1}^{\mathcal{S}}\tau^{-1}}{\sum_{\tau=1}^{\mathcal{S}} \tau^{-2}} \sim b \ln \mathcal{S}
    \end{align}
    and thus by rescaling $T_\mathcal{S}$ by $\mathcal{S}^2 \ln \mathcal{S}$, we get that
    \begin{align}
    \label{eq:MargLifetimeDist}
        \mathbb{P}(T_\mathcal{S}/(\mathcal{S}^2\ln \mathcal{S}) \geq t)=\mathbb{P}\left(N_\mathcal{S} \geq t \mathcal{S}^2/b \right)=\Xi_1(\sqrt{b/t})
    \end{align}
    which results  in a scaling form for the distribution of $T_\mathcal{S}$. 
\end{itemize}

To sum up, with Eqs.~\eqref{eq:RecLifetimeDist}, \eqref{eq:TransLifetimeDist} and \eqref{eq:MargLifetimeDist}, we have characterized  the asymptotic distributions (which obey a single-parameter scaling) of the lifetime of  starving RWs, from which the first two cumulants (up to log corrections) are given by
\begin{align}
\label{eq:TSmom}
    \left\langle T_{\mathcal{S}} \right\rangle, \; \sqrt{\text{Var}\left( T_{\mathcal{S}} \right)} \propto 
    \begin{cases}
        \mathcal{S}  &\mbox{ for } \mu <1 \; , \\
        \mathcal{S}^2 & \mbox{ for } \mu =1 \; , \\
        \exp\left[ \left( \mathcal{S}/a \right)^{\mu/(1+\mu)}\right] & \mbox{ for } \mu>1.
    \end{cases}
\end{align}

\begin{figure}
    \centering
    \includegraphics[width=0.5\columnwidth]{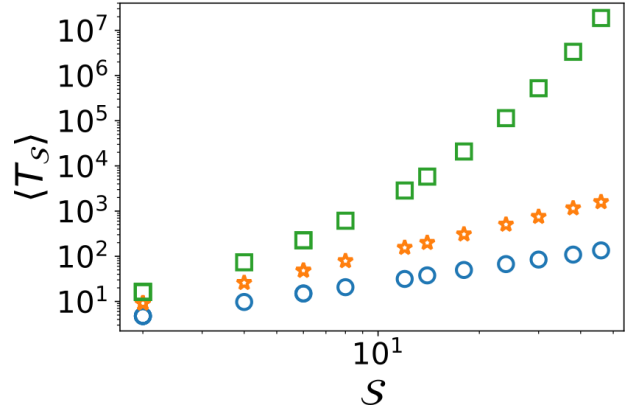}
    \caption{\leo{{\bf Comparison of the average lifetime for a nearest neighbor RW in different dimensions.} Blue circles stand for the $1d$ RW, orange squares the $2d$ RW and green squares the $3d$ RW.}}
    \label{fig:Lifetime}
\end{figure}

\leo{These results are particularly striking in the context of the nearest neighbor RW on the hypercubic lattice of dimensions $1$, $2$ and $3$ (see Fig.~\ref{fig:Lifetime}): the lifetime grows in radically different manner despite the fact that the RW moves diffusely.}

\section{Distribution of the position at starvation $R_\mathcal{S}$}

\begin{figure}[th!]
    \centering
    \includegraphics[width=\columnwidth]{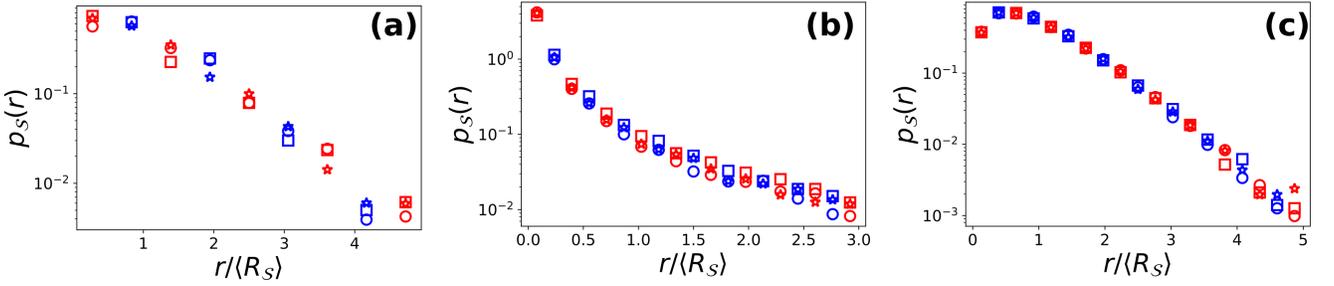}
    \caption{{\bf Check of the effective independence between position at starvation and lifetime.}\\ {\bf (a)}  RWs on the Sierpinski lattice (recurrent, $\mu=\ln 3/\ln 5$, $\mathcal{S}=10$, $100$ and $1000$), {\bf (b)} $1d$ L\'evy flights of parameter $\alpha=1$ (marginal, $\mu=1$, $\mathcal{S}=10$, $100$ and $1000$),  {\bf (c)} nearest neighbour jumps in $3d$ (transient, $\mu=3/2$, $\mathcal{S}=10$, $20$ and $40$). Independent simulations are in blue, the dependent ones are in red. Increasing values of $\mathcal{S}$ are represented via circles, stars and squares respectively.  }
    \label{fig:pos}
\end{figure}

\begin{figure}[th!]
    \centering
    \includegraphics[width=\columnwidth]{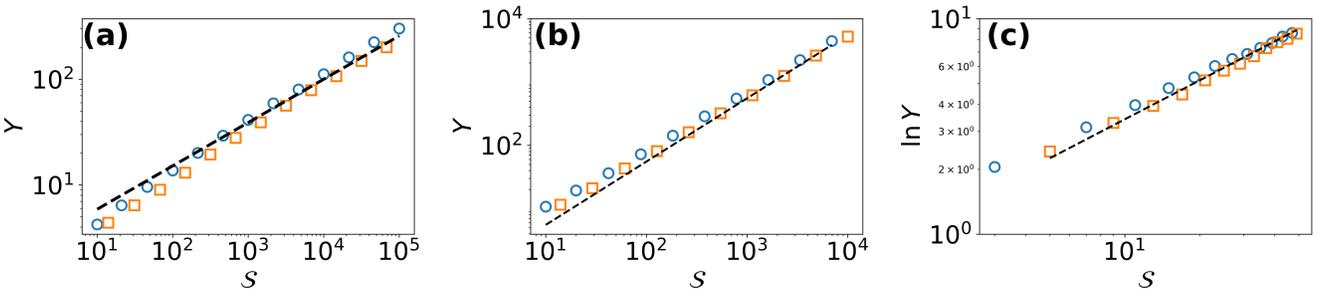}
    \caption{{\bf Moments of the position $R_\mathcal{S}$ at starvation.}\\ Average (blue circles) and standard deviation (orange squares) of $R_\mathcal{S}$ for {\bf (a)}  RWs on the percolation cluster (recurrent, $\mu\approx 0.659$ and $d_\text{w}\approx 2.44$ \cite{majid1984exact}), {\bf (b)} $2d$ nearest neighbour RWs (marginal, $\mu=1$ and $d_\text{w}=2$),  {\bf (c)} $3d$ nearest neighbour RWs (transient, $\mu=3/2$ and $d_\text{w}=2$). The black dashed lines correspond to Eq.~\eqref{eq:momR}. }
    \label{fig:Rmom}
\end{figure}

The distribution of the lifetime $T_\mathcal{S}$ allows to determine the position of the walker at starvation. Neglecting the correlation between the position and the lifetime (we test this approximation numerically in \ref{fig:pos}), we link the distribution of the position at starvation $\vec{R}_\mathcal{S}=\vec{r}(T_\mathcal{S})$ with that of the lifetime and the propagator $p_t(\vec{r})$ of the RW:
\begin{align}
    \mathbb{P}(\vec{R}_\mathcal{S}=\vec{r})&\approx \int_0^\infty {\rm d} t p_t(\vec{r})\mathbb{P}(T_\mathcal{S}=t).
\end{align}
For scale-invariant RWs the propagator takes the following form \cite{ben2000}:
\begin{align}
    p_t(\vec{r})=\frac{1}{t^\mu}\Pi \left( \frac{|\vec{r}|^{d_\text{w}}}{t} \right).
\end{align}
Taking into account the asymptotic scaling form for $T_\mathcal{S}$ of Sec.~\ref{sec:lifetime}, we deduce that of $R_\mathcal{S}=|\vec{R}_\mathcal{S}|$ by rescaling it with $\left\langle T_\mathcal{S}\right\rangle^{1/d_\text{w}}$:
\begin{align}
        \mathbb{P}(R_\mathcal{S}/\left\langle T_\mathcal{S}\right\rangle^{1/d_\text{w}}=r)&=\int_0^\infty {\rm d }t p_t(r \left\langle T_\mathcal{S}\right\rangle^{1/d_\text{w}} )\mathbb{P}(T_\mathcal{S}=t) \\
        &=\int_0^\infty  \frac{{\rm d }t}{t^\mu}\Pi \left( \frac{r^{d_\text{w}}\left\langle T_\mathcal{S}\right\rangle}{t} \right)\mathbb{P}(T_\mathcal{S}=t) \\
        &=\int_0^\infty  \frac{{\rm d }u}{u^\mu}\Pi \left( \frac{r^{d_\text{w}}}{u} \right)\frac{1}{\left\langle T_\mathcal{S}\right\rangle}\mathbb{P}(T_\mathcal{S}/\left\langle T_\mathcal{S}\right\rangle=u) \label{eq:pRS}
    \end{align}
which is independent of $\mathcal{S}$. With this, based on Eq.~\eqref{eq:TSmom}, we find that (see numerical check in \ref{fig:Rmom}),

\begin{align}\label{eq:momR}
    \left\langle R_\mathcal{S}\right\rangle  ,\sqrt{\text{Var}(R_\mathcal{S})} \propto
\begin{cases}
\mathcal{S}^{1/d_\text{w}} &  \mu<1\\
\mathcal{S}^{2/d_\text{w}}      & \mu=1\\
\exp\left[\frac{1}{d_\text{w}}\left( \frac{\mathcal{S}}{a} \right)^{\frac{\mu}{1+\mu}} \right]                       & \mu>1.
\end{cases}
\end{align}

Of note, due to the convolution in Eq.~\eqref{eq:pRS}, the large $R_\mathcal{S}$ behaviour of the distribution of the distance to the origin at starvation depends crucially on the form of the propagator's scaling function $\Pi$: it is algebraic for L\'evy flights \cite{Kolmogorov1954} and stretched exponential for a large class of random walks on fractals \cite{Barlow1988,bunde1990multifractal,jones1996transition} as can be seen in \ref{fig:pos}. For the simple 2d RW with nearest neighbour jumps, studied in \cite{Benichou_2016}, it is Gaussian, and we have at large $r$  (using Eqs.~\eqref{eq:MargLifetimeDist}, \eqref{eq:NSMarg} and a saddle-point method):
\begin{align}
        \mathbb{P}(R_\mathcal{S}/\left\langle T_\mathcal{S}\right\rangle^{1/2}=r)&=\int_0^\infty \frac{{\rm d }u}{2\pi u} \exp \left[-r^2/u \right] \mathbb{P}(T_\mathcal{S}/\left\langle T_\mathcal{S}\right\rangle=u)\\
        &\propto \int_0^\infty \frac{{\rm d }u}{u} \exp \left[-r^2/u -u\right]  \\
        &\propto \exp\left[-\text{const. }r \right]
\end{align}
where the constant is the minimum of $U(v)=1/v+v$ on the positive axis. This explains the exponential tail distribution of the rescaled position observed in \cite{Benichou_2016}. Also, $R_\mathcal{S}\propto \left\langle T_\mathcal{S}\right\rangle^{1/2}\sim \mathcal{S}$ as was observed numerically in \cite{Benichou_2016}. 

We conclude that the knowledge of the lifetime of the starving RW provides the asymptotic distribution of the position at starvation.

\section{Strongly inhomogeneous graphs}
\begin{figure}
    \centering
    \includegraphics[width=\columnwidth]{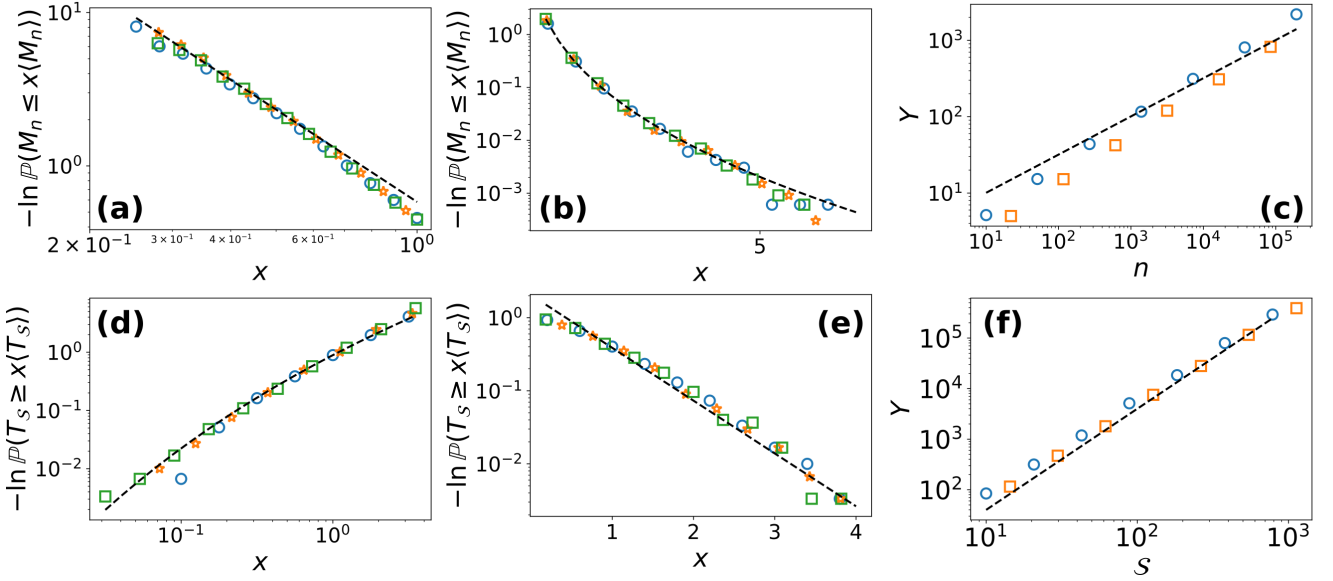}
    \caption{\leo{{\bf Distribution of the maximum and lifetime for a starving RW on a $(2,2)$ flower.} \textbf{(a)-(b)} Distributions of the maximum inter-visit time $M_n$ as a function of the rescaled variable $x=M_n/\left\langle M_n \right\rangle$ at respectively small and large $x$ for a RW on a $(2,2)$ flower of generation $10$ for $n=7196$, $37275$ and $193069$. \textbf{(c)} Average (blue circles) and standard deviation (orange squares) of $M_n$. \textbf{(d)-(e)} Distributions of the lifetime $T_\mathcal{S}$ as a function of the rescaled variable $x=T_\mathcal{S}/\left\langle T_\mathcal{S} \right\rangle$ at respectively small and large $x$ for $\mathcal{S}=183$, $379$ and $784$. \textbf{(f)} Average (blue circles) and standard deviation (orange squares) of $T_\mathcal{S}$. Increasing values of the parameters are in blue circles, orange stars and green squares. In all plots, the black dashed lines correspond to the best fit of the proportionality prefactors and process-dependent constants of the asymptotic expression of Eqs.~\eqref{eq:MargDistTn} and Eqs. (2) and (3) of the main text.  }}
    \label{fig:flower}
\end{figure}

\leo{We further illustrate our results by considering graphs with strong inhomogeneity, as is the case for the $(u,v)$ flowers defined in \cite{Meyer:2012,Rozenfeld:2007}. To obtain these graphs, one starts with a graph with one edge, and at each iteration every edge of the graph is replaced by end-connected two parallel chains of $u$ and $v$ edges (see Fig.2 (b) of \cite{Rozenfeld:2007}). In the case where $u>1$, it results in a graph of walk dimension $d_\text{w}=\frac{\log u v}{\log u}$ and $d_\text{f}=\frac{\log u+v}{\ln u}$ such that $\mu=\frac{\log u+v}{\log uv}$. All our results on the maximum $M_n$ \eqref{eq:MargDistTn} or the lifetime $T_\mathcal{S}$ \eqref{eq:MargLifetimeDist} still apply in this context, as we show for the RW on the $(2,2)$ flower which is a marginal ($\mu=1$) RW in \ref{fig:flower}.}

\section{A smooth process: the Random Acceleration Process}
\begin{figure}
    \centering
    \includegraphics[width=\columnwidth]{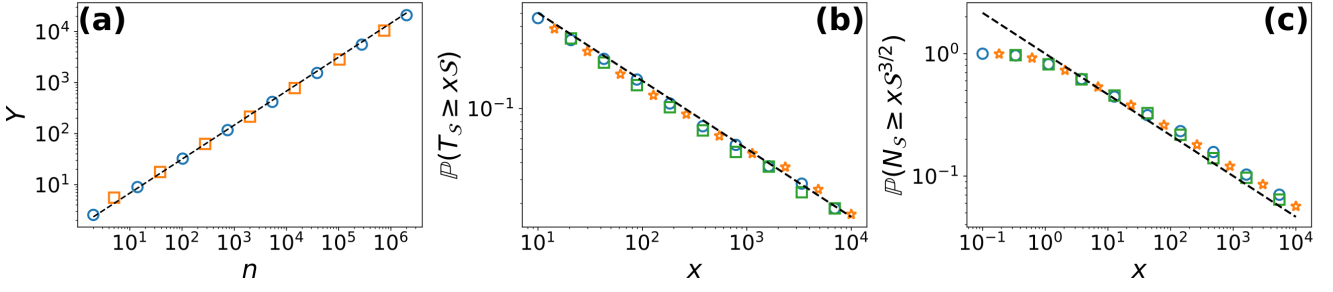}
    \caption{\leo{{\bf Maximum inter-visit time, lifetime and number of sites visited for a starving Random Acceleration Process.} \textbf{(a)} Average (blue circles) and variance (orange squares) of the maximum inter-visit time $M_n$. The black dashed line stands for $n^{2/3}$. \textbf{(b)} Distributions of the lifetime $T_\mathcal{S}$ as a function of the rescaled variable $x=T_\mathcal{S}/ \mathcal{S} $ at respectively small and large $x$ for $\mathcal{S}=60$, $80$ and $100$. The black dashed line is proportional to $x^{-1/2}$ \textbf{(c)} Distributions of the number of sites visited at starvation $N_\mathcal{S}$ as a function of the rescaled variable $x=N_\mathcal{S}/ \mathcal{S}^{3/2} $ at respectively small and large $x$ for $\mathcal{S}=60$, $80$ and $100$. The black dashed line is proportional to $x^{-1/3}$.}}
    \label{fig:StarRAP}
\end{figure}

\leo{As outlined in \cite{regnier2023record}, we expect that the correlations between the time to visit new sites will dominate for smooth processes (continuous processes with $d_\text{w}<1$), such as the Random Acceleration Process (RAP) \cite{bray2013persistence,Godreche:2021}. In this section, we display numerical results on the RAP and show how the results differ from the non-smooth case (in \ref{fig:StarRAP}, we consider that the RW visits all the sites it passes by). It is found that the distribution at large number of sites visited and long lifetimes are not exponential as was the case when the $\tau_k$ are independent, but algebraic (\ref{fig:StarRAP} \textbf{(b)} and \textbf{(c)}).}\\

\leo{ However, interestingly, we observe that the scaling of the maximum inter-visit times $M_n$, the number of sites visited at starvation $N_\mathcal{S}$ and the lifetime $T_\mathcal{S}$ are the same as the ones described in Eqs. (2), (3), (13) and (15) (see \ref{fig:StarRAP}). }

%